\documentclass[final]{siamltex}

\usepackage{graphicx}
\usepackage{ stmaryrd }

\usepackage{amsmath,amssymb} 
\usepackage[usenames,dvipsnames]{color}

\newcommand\xx{\mathbf{x}}
\newcommand\kk{\mathbf{k}}
\newcommand\xtt{\mathbf{h}}

\newcommand{\FF}{\mathbf{F}}

\newcommand{\eg}{\emph{e.g.} }
\newcommand{\ie}{\emph{i.e. }}

\def\tvtr{\mbox{$\mathbf{{u}}$}}
\def\wwr{\mbox{$\mathbf{{v}}$}}
\def\ww{\mbox{$\mathbf{{v}}$}}
\def\tvt{\mbox{$\mathbf{{u}}$}}

\newtheorem{algo}{Algorithm}
\newtheorem{remark}{\it Remark}

\title{ Self-similar prior and wavelet bases for hidden incompressible turbulent motion}

\author{P. H\'EAS\thanks{Inria Rennes, Centre de Bretagne Atlantique, Universit\'e de Beaulieu, 35042 Rennes, France ({\tt Patrick.Heas@inria.fr}) } \and   F. LAVANCIER\thanks{Universit\'e de Nantes,
Laboratoire de Math\'ematiques Jean Leray, 2 rue de la Houssini\`ere, 44322 Nantes, France}   \and S. KADRI-HAROUNA\thanks{Universit\'e de La Rochelle, 23 avenue Albert Einstein, 17071 La Rochelle, France  }
   }

\begin{document}

\maketitle

\begin{abstract}
This work is concerned with the ill-posed inverse problem of estimating turbulent flows from the observation of an image sequence. From a Bayesian perspective, a divergence-free isotropic fractional Brownian motion (fBm) is chosen as a  prior model for instantaneous turbulent velocity fields.
This self-similar prior characterizes accurately second-order statistics of velocity fields in incompressible isotropic turbulence. 
Nevertheless, the associated maximum a posteriori involves a fractional Laplacian operator which is delicate to implement in practice.
To deal with this issue, we propose to decompose the divergence-free fBm on well-chosen wavelet bases.  As a first alternative, we propose to design wavelets as whitening filters. We show that these filters are fractional Laplacian wavelets composed with the Leray projector.
As a second alternative, we  use a divergence-free wavelet basis, which takes implicitly into account the incompressibility constraint arising from physics. Although the latter decomposition involves correlated wavelet coefficients, we are able to handle this dependence in practice.
Based on these two wavelet decompositions, we finally provide effective and efficient algorithms to approach the maximum a posteriori. An intensive numerical evaluation proves the relevance of the proposed wavelet-based self-similar priors.
 \end{abstract}

\begin{keywords} 
Bayesian estimation, fractional Brownian motion, divergence-free wavelets, fractional Laplacian, connection coefficients, fast transforms, optic-flow,   isotropic turbulence. 
\end{keywords}

\begin{AMS}
60G18, 60G22, 60H05, 62F15, 65T50, 65T60 
\end{AMS}

\pagestyle{myheadings}
\thispagestyle{plain}
\markboth{}{ Self-similar  prior and wavelet bases for hidden incompressible turbulent motion}

\section{Introduction}
 
This work is concerned with the ill-posed inverse problem of estimating turbulent motions from the observation of an image sequence. Turbulence motion phenomena are  often studied in the context of incompressible fluids, which is the setting of this paper.  This inverse problem  arises in the context of experimental physical  settings, where one is interested in recovering the kinematical state of an incompressible turbulent fluid flow from  the observation of a  sequence of images, \eg particle image velocimetry in experimental fluid mechanics, wind or ocean currents retrieval from satellite imagery in geophysics.  Solving accurately this type of inverse problems constitute an important issue since a complete physical theory is still missing for  turbulence phenomenology. 

More specifically, the above inverse problem can be viewed as a Maximum A Posteriori (MAP) estimation of a vectorial field  $\mathbf{u}$ over a space of admissible solutions: 
$$\mathbf{u}^*({y}_0, {y}_1 ) \in \arg\max_{\mathbf{u}} \,p_{\delta {y} | \mathbf{u}} p_{\mathbf{u}},$$
where ${y}_0$ and ${y}_1$ are two consecutive observed images, $\mathbf u$  is a velocity field, and $\delta y$ is a function of $y_0$, ${y}_1$ and $\mathbf u$, which characterizes the evolution between ${y}_0$ and ${y}_1$. In this problem, $\mathbf u$ is not observed and the incompressibility constraint demands the motion field $\mathbf u$ to be divergence-free.
In this Bayesian framework, $p_{\delta y| \mathbf{u}}$ denotes the likelihood model, which relates
the motion  of the physical system to the spatial and temporal variations of the image intensity. The adjunction of  a prior information $p_{\mathbf{u}}$ for the velocity field $\mathbf u$ is in this case mandatory since, as we will see in section \ref{sec:Bayes}, this non-linear problem is  under-constrained.

Concerning the choice of  $p_{\delta y| \mathbf{u}}$, relevant physical  models have been proposed for fluid flow imagery. A review  in the context of experimental fluid mechanics  can be found in  \cite{Liu08}. We will assume in this paper the simple model where ${y}_0$ and ${y}_1$ are the solutions between two consecutive times of a transport equation driven by $\mathbf u$, see section \ref{sec:Bayes} for more details.

Concerning the choice of $p_{\mathbf{u}}$, the incompressible Navier-Stokes equations perfectly describe the structure of an incompressible velocity field, \ie  the prior for $\mathbf{u}$. However, this implicit choice of constraints leads to an  optimization problem which is often severely  ill-conditioned and  computationally prohibitive. Some recent  works propose to use a simplified version of Navier-Stokes equations to circumvent this issue (see \eg \cite{Corpetti09,Papadakis08}).  On the other hand, instead of relying the prior on the Navier-Stokes equations, spatial regularizers of $\mathbf{u}$ have been proposed to serve as a prior. A first approach in this direction is to assume a low-dimensional parametric form for  $\mathbf{u}$, see \eg  \cite{Becker11, Derian12wavelet}. A second strategy is to choose a prior that introduces some spatial smoothness for  $\mathbf{u}$. Typically, the regularisation penalises in this case the norm of the gradient (or higher derivatives) of  $\mathbf{u}$,  see \eg \cite{Horn81,KadriHarouna12, Suter94,Yuan07}. Finally, a third approach consists in the introduction of a self-similar constraint on $\mathbf{u}$.  Self-similarity is a well-known feature of turbulence,  theoretically and experimentally attested, see \eg\cite{Monin07}. An attempt in this direction has been conducted  in \cite{Heas12c,Heas11a}. 
Besides, a general family of self-similar regularizers has been introduced in \cite{Tafti11}. In the same spirit, our choice for the prior of $\mathbf{u}$ is the divergence-free isotropic fractional Brownian motion (fBm), as we now justify.

In addition to self-similarity, we assume $\mathbf{u}$ to be Gaussian. Non-Gaussian turbulent fields is an interesting alternative, which could potentially  describe more accurately the structure of turbulence \cite{Frisch95}, but they will not been considered in this paper. Note that the definition of such models is still an active domain of investigation, see \cite{Chevillard10,Kolmogorov41,Kraichnan67,Monin07,RobertVargas08}. Assuming further the stationarity of the increments of $\mathbf{u}$ leads necessarily to $\mathbf{u}$ being a vector fBm (see \cite{Tafti10}). In order to satisfy the incompressibility constraint, $\mathbf{u}$ is finally demanded to be divergence-free. Although divergence-free fBm's can be viewed as a limiting case of the vector fBm's considered in  \cite{Tafti10}, we provide  a  proper definition   in section \ref{app:spectralDef}. In particular, a spectral integral representation is deduced. 
 
This choice of prior involves in practice   fractional Laplacian operators, which are numerically delicate to implement. Indeed, there is a lack  of  effective algorithms in the literature able to deal in practice with those particular priors. In  \cite{Tafti11}, the authors circumvent this issue in their numerical applications by limiting themselves  to  non-fractional settings. To tackle this problem, we  propose to decompose  fBms on well-chosen  wavelet bases. This strategy allows us to
expand fractional Laplacian operators on the wavelet components and makes their computation feasible.  We focus on two particular wavelet bases.
\begin{itemize}
\item[-] As a first alternative, we propose to design  wavelets as whitening filters for divergence-free isotropic fBms. For scalar fBms indexed by time, fractional  wavelet bases represent  ideal whitening filters \cite{Majda97,Meyer99}.   In \cite{Tafti09,Tafti09b,Tafti10}, the authors extend these bases to the case of isotropic vector fields, namely they use the so-called fractional Laplacian polyharmonic spline wavelets.  In our case of  divergence-free isotropic vector fBms, we  show that for any mother wavelet,  fractional Laplacian wavelet series composed with the Leray projector  is an appropriate whitening filter.

\item[-]  The second alternative is to use divergence-free wavelet bases, which are well-suited to our case. These wavelets simplify the decomposition, since  they do not involve fractional operators and make superfluous the use of  Leray projector \cite{ Deriaz06,Deriaz08, KadPer12,Lemariet92}.  However, the wavelets coefficients are then correlated. 
\end{itemize}
As seen in section \ref{sec:optim_frac},   implementation for the first bases can be accurately performed in the Fourier domain. For the second bases, the computation of the associated posterior density relies on the covariances of the wavelet coefficients. We provide a closed form expression for  these covariances, which allows us to  propose an approximation of the posterior using  wavelet connection coefficients, see sections \ref{sec:SecDiv0Optim} and \ref{sec:SpatialOptim}.

Moreover, an additional pleasant feature of wavelet representations is that it is well adapted to the non-convex optimization  problem of  MAP estimation, since it naturally provides a multi-resolution approach. As a matter of fact, this approach  has proved to be experimentally efficient for motion estimation problems (subjected to different priors)   \cite{Derian12wavelet, KadriHarouna12, Wu00}.  The optimization algorithms   rely on Fast Fourier Transforms (FFT) or on Fast Wavelet Transforms (FWT).  Finally, for the divergence-free wavelet  decomposition, we introduce an approximated MAP optimization  procedure, which turns out to be by far the fastest   algorithm while being as accurate as the other approaches. 

A numerical comparison with state-of-the-art procedures is performed in section~\ref{sec:results} on a general benchmark of divergence-free fBm. Note that, while these fields  match perfectly our priors, they are likely to be in agreement with real fluid flows according to turbulence phenomenology, as mentioned before. As a result, our procedure seems to recover more accurately the hidden motion field, in particular when the Hurst parameter $H$ corresponds to 2D and 3D turbulence models, respectively $H=1$ and $H=\frac13$ according to physics.

The paper is organized as follows. In section \ref{app:spectralDef}, we give a spectral representation  of  divergence-free isotropic fBms, which will serve as the basic definition all along the paper. In section \ref{sec:wav}, we then provide the two wavelet representations described above. Section~\ref{sec:Bayes} displays the Bayesian modeling of turbulent flows in terms of wavelet coefficients and considers the associated MAP estimators. In section~\ref{sec:optim}, gradient descent optimization algorithms achieving MAP estimation are presented for both wavelet bases. The numerical evaluation of the proposed regularizers is presented in section~\ref{sec:results}. Finally, the appendix gathers the technical proofs and some details about the algorithms and the computation of fractional Laplacian wavelet connection coefficients.

\section{Spectral definition of  divergence-free isotropic fBm}\label{app:spectralDef}

The unique self-similar  zero-mean Gaussian  process with stationary increments is the fractional Brownian motion (fBm), introduced in \cite{Mandelbrot68}. The definition of scalar fBm indexed by time has been extended: to the case of multi-dimensional state spaces, \ie {\it vector} processes indexed by time \cite{Amblard10}; to the {\it field} case, namely  scalar {\it isotropic fields} \cite{Reed95};  and to both cases, \ie {\it isotropic vector fields} \cite{Monin07,Tafti10}.

As far as we are interested in the construction of a prior for turbulent vector fields, we are particularly concerned with \emph{divergence-free} isotropic vector fBm in the following sense. 
For sake of clarity,  this work is restricted to  the bi-variate case although it is possible to extend  our results to higher dimensions.  \\

\begin{definition}\label{def:fbm}
A bi-variate field $\tvt(\xx)\triangleq (\tvt_1(\xx),\tvt_2(\xx))^T$, $\xx\in\mathbb R^2$,  is a divergence-free isotropic bi-variate fBm with parameter $0<H<1$ if 
\begin{itemize}
\item $\tvt$ is Gaussian;
\item $\tvt$ is self-similar, \ie for any $\lambda>0$, $\tvt(\lambda\xx) \overset{ \mathcal{L}}{=}\lambda^H\tvt(\xx)$; 
\item $\tvt$ has stationary  increments, \ie $\forall \mathbf h \in\mathbb R^2$, $\tvt(\xx)-\tvt(\xx-\mathbf h)$ is stationary; 
\item $\tvt$ is isotropic, \ie for any rotation matrix  $M$,  $\tvt( \xx)  \overset{ \mathcal{L}}{=} \tvt(M\xx)$ ; 
\item $\tvt$ is  divergence-free, \ie $\emph{div}\, \tvt=0$ almost surely.
\end{itemize}
\end{definition}
%

Since  a fBm is  almost surely not differentiable, the divergence operator in the last property is to be understood in a weak sense:  for any  test function $\psi \in \mathcal{C}^1(\mathbb{R}^2)$ with a fast decay at infinity, 
\begin{align*}
\textrm{div}\, \tvt&\triangleq\langle \textrm{div}\, \tvt , \psi \rangle =- \langle \tvt / \nabla \psi \rangle,
\end{align*}  
 where $\langle .,.\rangle$   (resp.  $\langle ./. \rangle$) denotes   the inner product of two scalar (resp. two  bi-variate) functions, and $\nabla$ denotes the gradient operator.  
Let us remark  that differentiability in a weak sense is a little bit schematic when compared to real flows, since it is well known that below the Kolmogorov scale,  the field becomes smooth. The fBm model only constitutes an approximation of turbulence, consistent with the range of  scales modelled in practice.

%

Isotropic fractional Brownian vector fields have been introduced by Tafti and Unser \cite{Tafti10} as the solution of a fractional Poisson equation, or \emph{whitening} equation, see \cite{Tafti09b} and \cite{Tafti10}. Isotropic \emph{divergence-free} fractional Brownian vector fields, as we require, turns out to be  a limiting case of the solution (corresponding to $\xi_{irr}=\infty$ and $\xi_{sol}=0$ in the setting of \cite{Tafti10}). Specifically, it can be defined by:
\begin{equation}\label{sol_poisson} \tvt= \sigma (-\acute{\Delta})_{div}^{-\frac{H+1}{2}}\bf W,\end{equation}
where $H\in(0,1)$ is the Hurst parameter, $\sigma$ is a positive constant, $\bf W$ is a vector of two independent Gaussian white noise, and $(-\acute{\Delta})_{div}^{-\frac{H+1}{2}}$ is a fractional Laplacian operator, defined for any $f$ in $(L^2(\mathbb R^2))^2$ by
{\color{black}
$$(-\acute{\Delta})_{div}^{-\frac{H+1}{2}} f (\xx) = \frac 1 {(2\pi)^2} \int_{\mathbb R^2}  \frac{(e^{i{\bf k} \cdot \xx} -1)}{\|\kk\|^{H+1}}   \left[ \mathbf{I} - \frac{\kk\kk^T}{ \| {\kk }\|^2}\right]  \mathcal F(f) (\kk) d\kk,$$
where $\mathcal F(f)$ denotes the Fourier transform of $f=(f_1,f_2)^T$ in $(L^2(\mathbb R^2))^2$, viz. $$\mathcal F(f)(\kk) = \left(\int_{\mathbb{R}^2}  f_1(\xx) e^{-i\kk \cdot \xx} d\xx,\int_{\mathbb{R}^2}  f_2(\xx) e^{-i\kk \cdot \xx} d\xx\right)^T.$$
}

In the following proposition, we provide a spectral representation of  {\color{black} $\tvt$}    defined by \eqref{sol_poisson}. This proposition gives an explicit representation, therefore it can also be considered as an alternative definition. In fact, we will only refer in the sequel to this representation when we consider isotropic divergence-free fractional Brownian vector fields.\\

\begin{proposition} \label{prop:spectral}
The  isotropic divergence-free fractional Brownian vector field $\tvt$, as defined by \eqref{sol_poisson} admits the following representation, for any  $H\in(0,1)$,
\begin{equation}\label{spectral}
 \tvt (\xx) = \frac{\sigma}{2\pi}\int_{\mathbb{R}^2} \frac{(e^{i{\bf k} \cdot \xx} -1)}{\|\kk\|^{H+1}}   \left[ \mathbf{I} - \frac{\kk\kk^T}{ \| {\kk }\|^2}\right]  \tilde {\bf W} (d {\bf k}),\,\,  \xx \in \mathbb{R}^2,
\end{equation}
where $\mathbf{I}$ denotes the identity matrix and  $\tilde {\bf W}=(\tilde W_1,\tilde W_2)^T$ denotes a bi-variate standard Gaussian spectral measure, \ie $\tilde W_1$ and $\tilde W_2$ are independent  and for $i=1,2$, for any Borel sets $A, B$ in $\mathbb R^2$,  $\tilde W_i(A)$ is a centered complex Gaussian random variable,  $\tilde W_i(A)=\overline{\tilde W_i(-A)}$ and $\mathbb{E}(\tilde W_i(A)\overline{\tilde W_i(B)})=|A\cap B|$.\\

\end{proposition}

As a by-product, we deduce the following structure matrix function that characterizes the law of the Gaussian vector field $\tvt$: for any $i,j=1,2$, for any $\xx_1,\xx_2,\xx_3,\xx_4$ in $\mathbb R^2$, 
\begin{multline}\label{eq:crosscorrel} 
{\bf \Sigma}_{ij}(\xx_1,\xx_2,\xx_3,\xx_4)\triangleq \mathbb{E}[(\tvt_i(\xx_2)-\tvt_i(\xx_1))\overline{(\tvt_j(\xx_4)-\tvt_j(\xx_3))}]\\= \sigma^2 c_H\left(F^H_{ij}(\xx_2-\xx_3)-F^H_{ij}(\xx_2-\xx_4) - F^H_{ij}(\xx_1-\xx_3)+F^H_{ij}(\xx_1-\xx_4)\right),
\end{multline}
where $c_H={\Gamma(1-H)}/(\pi 2^{2H+2} \Gamma(H+1) H(2H+2))$ {\color{black} with $\Gamma(.)$ denoting the Gamma function,} and $$F^H(\xx)=\|\xx\|^{2H} \left((2H+1)\mathbf{I} - 2H \frac{\xx\xx^T}{\|\xx\|^2}  \right).$$
In particular,  taking $\xx_1=\xx_2-\xtt$ and $\xx_3=\xx_4-\xtt$, for some $\xtt\in\mathbb R^2$, shows  the  power-law structure for the second order moment of the increments. We can also deduce the generalized power spectrum $E_j$, $j=1,2$, of each component $\tvt_j$ of  $\tvt$, defined implicitly by (see \cite{Flandrin89, Reed95}):
\begin{multline*}{\bf \Sigma}_{jj}(\xx_1,\xx_2,\xx_3,\xx_4) \\ = \frac{1}{(2\pi)^2} \int_{\mathbb{R}^2}\ E_j(\kk) \left(e^{i\kk \cdot (\xx_2-\xx_4)}-e^{i\kk \cdot (\xx_2-\xx_3)}-e^{i\kk \cdot (\xx_1-\xx_4)}+e^{i\kk \cdot (\xx_1-\xx_3)}\right)d\kk.\end{multline*}
We deduce from the proof in Appendix \ref{proofofprop:spectral}:
\begin{equation}\label{powerspectrum} E_j (\kk) = \sigma^2 \left(1-\frac{k_j^2}{\|\kk\|^2}\right)\|\kk\|^{-2H-2}.\end{equation}


While all properties required in Definition \ref{def:fbm} can be found in \cite{Tafti10} going back to the definition \eqref{sol_poisson}, they are straightforward consequences of the spectral representation \eqref{spectral}:  Gaussianity and $H$-self-similarity of $\tvt$ are easily seen from \eqref{spectral}; Stationarity of the increments and isotropy follow from \eqref{eq:crosscorrel};  Finally the divergence-free property of $\tvt$ is a consequence of the presence of the Leray projection operator  $\left[ \mathbf{I} - \frac{\kk\kk^T}{ \| {\kk }\|^2}\right]$ in \eqref{spectral}, and will appear clearly in the wavelet decomposition of $\tvt$ considered in the next section.\\

\begin{remark}\label{rem1}
The definition of $\tvt$ in \eqref{sol_poisson} can be extended to $H>1$, see \cite{Tafti10}. Similarly, the spectral representation \eqref{spectral} can be extended to $H>1$ by application of successive integrations as in \cite{Reed95}, leading to the representation 
\begin{equation}\label{spectralgeneral}\tvt (\xx) = \frac{\sigma}{2\pi}\int_{\mathbb{R}^2} \frac{(e^{i{\bf k} \cdot \xx}  - \sum_{j=0}^{\lfloor H\rfloor}(i{\bf k} \cdot \xx)^j/j!  )}{\|\kk\|^{H+1}}   \left[ \mathbf{I} - \frac{\kk\kk^T}{ \| {\kk }\|^2}\right]  \tilde {\bf W} (d {\bf k}).\end{equation}
In this case $\tvt$ has stationary increments of order $N=\lfloor H\rfloor +1$, in the sense that for any $({\bf h}_N,\dots,{\bf h}_1)\in(\mathbb R^2)^N$, the successive symmetric differences $D_{{\bf h}_N}\dots D_{{\bf h}_1}\tvt(\xx)$ form a stationary process, where $D_{{\bf h}}: {\bf f}(\cdot) \rightarrow {\bf f} (\cdot+\frac{{\bf h}}{2})-{\bf f}(\cdot-\frac{{\bf h}}{2})$. These increments write specifically in this case:
$$D_{{\bf h}_N}\dots D_{{\bf h}_1}\tvt(\xx) =  \frac{\sigma}{2\pi}\int_{\mathbb{R}^2} e^{i{\bf k} \cdot \xx}  \prod_{j=1}^N 2i\sin\left(\frac{\kk\cdot {\bf h}_j}{2}\right)\, \|\kk\|^{-H-1}   \left[ \mathbf{I} - \frac{\kk\kk^T}{ \| {\kk }\|^2}\right]  \tilde {\bf W} (d {\bf k}).$$
\end{remark}

\section{Wavelet representations of divergence-free isotropic fBm}\label{sec:wav}

This section presents  wavelet expansions of the divergence-free isotropic fBm representation of Proposition~\ref{prop:spectral}.  As pointed out  in the introduction,  wavelet bases are chosen in order to make fractional calculus involved by fBms feasible and effective. 
In practice we observe $\tvt$ on a bounded domain, say $[0,1]^2$. For this reason, we focus  on the wavelet decomposition of $\tvt$ in $(L^2([0,1]^2))^2$. In the first part, the expansion of $\tvt$ relies on a fractional Laplacian wavelet  basis and finally involves uncorrelated bi-variate random coefficients.  In the second part, {\color{black}  the expansion relies on a divergence-free wavelets basis, which is well-adapted to our setting, and involves mono-variate but correlated random coefficients.}
These two wavelet expansions  will then allow us to  derive  efficient optimization algorithms  (see  section \ref{sec:optim})   solving the MAP estimation problem presented in the introduction.

The construction of these two decompositions relies on an orthonormal wavelet basis of $L^2([0,1])$. There are several methods to build  orthonormal  wavelet bases of $L^2([0,1])$   (see  \cite{mallat2008wavelet}, chapter 7). For ease of presentation, we consider the simplest one, which consists in periodizing scalar wavelets of $L^2(\mathbb R)$.
Let   $\psi$ be a mother wavelet  with compact support  and its associated wavelets dilated at scale $2^{-s+1}$ and translated by $2^{-s+1}\ell $:
\begin{equation}\label{defpsi} \psi_{\ell,s}(x)\triangleq 2^{(s-1)/2}\psi(2^{s-1}x-\ell).\end{equation}
The wavelet set $\{\psi_{\ell,s}(x); x\in\mathbb R, \ell, s \in \mathbb{Z}\}$ form an orthonormal basis of $L^2(\mathbb R)$.
 The periodized  wavelets $\psi^{per}_{\ell, s}$, $\ell, s \in \mathbb{Z}$  is then defined by 
\begin{equation}\label{psiper}
\psi^{per}_{\ell, s}(x) =\sum_{k=-\infty}^{k=+\infty}\psi_{\ell, s}(x+k), \quad x \in [0,1].\end{equation}
The set $\{\psi^{per}_{\ell,s}(x); x\in[0,1], s>0, 0\leq \ell<2^{s-1}\}$ with the indicator function over $[0,1]$ form an orthonormal basis of $L^2([0,1])$. We will assume in the sequel that $\psi$ has  $M>H$ vanishing moments and is {\color{black} $\max(H+2,\,2H)$ }  times differentiable. 

\subsection{Fractional Laplacian wavelet decomposition}\label{sec:wave}
Since a divergence-free fBm does not belong to $(L^2(\mathbb R^2))^2$ almost surely, it can not be in principle decomposed in a wavelet basis of this space, unless we resort to generalized random fields \cite{Hida04}. However, even in the later setting, this kind of decomposition typically involves a sum depending on arbitrarily  large scales, which is not suitable for application  (see  \cite{Meyer99} in the standard fBm case). In contrast, we show in Proposition~\ref{corrol01} below that in our case where $\tvt$  is considered on the compact domain $[0,1]^2$, and under a mild condition (namely $\int_{[0,1]^2} \tvt(\xx)d\xx=0$), the divergence-free fBm enjoys a simple tractable decomposition in $(L^2([0,1]^2))^2$ with respect to fractional Laplacian wavelets.






The bi-dimensional wavelet basis of $(L^2([0,1]^2))^2$  is  constructed from  $\psi^{per}_{\ell, s}$  as follows.
Denoting $\mathbb{I}_{A}$ the indicator function over the set $A$,  we form the three following wavelet sets: 
\begin{align*}
&\{ \Phi_{\ell_1, s_1,0,0}=\psi^{per}_{\ell_1, s_1}(x_1)\mathbb{I}_{[0,1]}(x_2);\; 0\!\le \!\ell_1\!<\!2^{s_1-1},  s_1> 0\},\\
&\{ \Phi_{0,0,\ell_2,s_2} =\mathbb{I}_{[0,1]}(x_1) \psi^{per}_{\ell_2, s_2}(x_2);\; 0\!\le \!\ell_2\!<\!2^{s_2-1},  s_2> 0\},\\
&\{  \Phi_{\ell_1, s_1,\ell_2,s_2} =\psi^{per}_{\ell_1, s_1}(x_1)\psi^{per}_{\ell_2, s_2}(x_2);\; 0\!\le\! \ell_1\!<\!2^{s_1-1}, 0\!\le \!\ell_2\!<\!2^{s_2-1},  s_1,s_2 > 0 \}.
\end{align*}
Let us denote by $\Omega$ the set of indices $({\boldsymbol \ell},{\boldsymbol s})=(\ell_1,s_1,\ell_2,s_2)$ involved in these three sets and $\{\Phi_{{\boldsymbol \ell},{\boldsymbol s}};{({\boldsymbol \ell},{\boldsymbol s})} \in \Omega\}$ the union of them. An orthonormal basis of $L^2([0,1]^2)$ is finally the union of the latter set with the indicator function $\mathbb{I}_{[0,1]^2}(\xx)$ (see Theorem 7.16 in  \cite{mallat2008wavelet}). A basis of the product space $(L^2([0,1]^2))^2$ is deduced by tensorial product.


\textcolor{black}{ Let us now consider the extension of ${\Phi}_{{\boldsymbol \ell},{\boldsymbol s}}$ to $L^2(\mathbb R^2)$ that vanishes outside $[0,1]^2$, which will be denoted by ${\Phi}^0_{{\boldsymbol \ell},{\boldsymbol s}}$ :
$${\Phi}^0_{{\boldsymbol \ell},{\boldsymbol s}}(\xx)=\begin{cases} {\Phi}_{{\boldsymbol \ell},{\boldsymbol s}}(\xx)&\text{if }\xx\in[0,1]^2,\\ 0&\text{if }\xx\notin[0,1]^2.\end{cases}$$
For any ${({\boldsymbol \ell},{\boldsymbol s})} \in \Omega$, we define the  fractional Laplacian wavelets, that  correspond to an integration of order $H+1$ of ${\Phi}^0_{{\boldsymbol \ell},{\boldsymbol s}}$:
\begin{align}\label{eq:fracdiff}{\Phi}^{(-H-1)}_{{\boldsymbol \ell},{\boldsymbol s}}(\xx) \triangleq \mathcal F^{-1}( \kk\mapsto \| {\bf k } \| ^{-H-1}\mathcal F({\Phi}^0_{{\boldsymbol \ell},{\boldsymbol s}})({\bf k})),\hspace{0.3cm} \forall {({\boldsymbol \ell},{\boldsymbol s})} \in \Omega,\end{align}
where $\mathcal F$ denotes the Fourier  operator in $L^2(\mathbb R^2)$ and $\mathcal F^{-1}$ the inverse Fourier operator,  
$$\mathcal F^{-1}(f)(\xx)= \frac{1}{(2\pi)^2}\int_{\mathbb{R}^2} e^{i{\bf k} \cdot \xx} f(\kk) d\kk,\quad \forall f\in L^2(\mathbb R^2).$$
}
This fractional integration operator can be denoted in the spatial domain by $(-\Delta)^{\frac{-H-1}{2}}$, following  \cite{Meyer99}, leading to the relation \textcolor{black}{${\Phi}^{(-H-1)}_{{\boldsymbol \ell},{\boldsymbol s}}(\xx)={(-\Delta)^{\frac{-H-1}{2}}{\Phi}^0_{{\boldsymbol \ell},{\boldsymbol s}}}({\xx})$}. Note that $\{\Phi^{(-H-1)}_{{\boldsymbol \ell},{\boldsymbol s}};{({\boldsymbol \ell},{\boldsymbol s})} \in \Omega\}$ constitutes a new family of wavelets, which is not orthogonal, unlike $\{\Phi_{{\boldsymbol \ell},{\boldsymbol s}};{({\boldsymbol \ell},{\boldsymbol s})} \in \Omega\}$, but biorthogonal, where ${\Phi}^{(H+1)}_{{\boldsymbol \ell},{\boldsymbol s}}$ is the dual wavelet of ${\Phi}^{(-H-1)}_{{\boldsymbol \ell},{\boldsymbol s}}$, \ie     $\langle {\Phi}^{(H+1)}_{{\boldsymbol \ell},{\boldsymbol s}},{\Phi}^{(-H-1)}_{{\boldsymbol \ell}',{\boldsymbol s}'}\rangle =\delta_{{\boldsymbol \ell},{\boldsymbol \ell}'}\delta_{{\boldsymbol s},{\boldsymbol s}'}$ for all ${{\boldsymbol \ell},{\boldsymbol \ell}'} ,{\boldsymbol s},{\boldsymbol s}' $ (see \cite{ Abry96,Blu02,Meyer99}).

 Let us finally recall the definition of the Leray projector, denoted by $\mathcal{P}$, that maps  square-integrable bi-variate   functions $\ww$ \textcolor{black}{in $(L^2(\mathbb R^2))^2$} onto the space of divergence-free functions:
{\color{black}
 \begin{align}\label{eq:ProjLerayDef}
 \mathcal{P} {\ww} \triangleq \mathcal F^{-1} \left(  \kk\mapsto \left[ \mathbf{I} - \frac{\kk\kk^T}{ \| {\kk }\|^2}\right]  \mathcal F(\ww)({\bf k})\right),
 \end{align}
 where for a bivariate function $\ww=(\ww_1,\ww_2)^T\in (L^2(\mathbb R^2))^2$, $\mathcal F \ww = (\mathcal F\ww_1,\mathcal F\ww_2)^T$ and similarly for $\mathcal F^{-1}$.}
This projection operator can be \textcolor{black}{formally} represented in the spatial domain by  \textcolor{black}{$\mathcal{P}({\ww}) =[\ww-\nabla \Delta^{-1}(\nabla \cdot \ww)]$}, for sufficiently smooth functions $\ww$.\\

We are now in position to present the wavelet decomposition of $\tvt$ in $(L^2([0,1]^2))^2$. We assume for simplicity that    $\int_{[0,1]^2} \tvt(\xx)d\xx=0$. Note that in practice, we observe $\tvt$ on a lattice with $n \times n$  sites, so the latter assumption roughly means that the mean value of $\tvt$ on this lattice is assumed to be negligible. 
 \\

\begin{proposition}\label{corrol01}
 Let $\tvt$ be an  isotropic divergence-free fBm with parameter $H \in (0,1)$.  Assuming $\int_{[0,1]^2} \tvt(\xx)d\xx=0$, we have in  $(L^2([0,1]^2))^2$:
\begin{align}\label{eq:WavSeries01}
\tvt(\xx) &=\sum_{({\boldsymbol \ell},{\boldsymbol s})\in \Omega}  \mathcal{P} \left[  \boldsymbol\epsilon_ {{\boldsymbol \ell},{\boldsymbol s}}{\Phi}^{(-H-1)}_{{\boldsymbol \ell},{\boldsymbol s}}\right]({\xx}),
\end{align} 
where {\color{black} coefficients $\boldsymbol\epsilon_ {{\boldsymbol \ell},{\boldsymbol s}}\triangleq (\epsilon^1_{{\boldsymbol \ell},{\boldsymbol s}},\epsilon^2_{{\boldsymbol \ell},{\boldsymbol s}})^T$ are  i.i.d. bi-variate  zero-mean  Gaussian random variables with variance $( 2\pi\sigma)^2\boldsymbol I$, and where $\Phi^{(-H-1)}_{{\boldsymbol \ell},{\boldsymbol s}}$, ${({\boldsymbol \ell},{\boldsymbol s})} \in \Omega$, are defined in \eqref{eq:fracdiff}, so that $\boldsymbol\epsilon_{{\boldsymbol \ell},{\boldsymbol s}} {\Phi}^{(-H-1)}_{{\boldsymbol \ell},{\boldsymbol s}}$ is the bivariate vector $(\epsilon^1_{{\boldsymbol \ell},{\boldsymbol s}}{\Phi}^{(-H-1)}_{{\boldsymbol \ell},{\boldsymbol s}},\epsilon^2_{{\boldsymbol \ell},{\boldsymbol s}}{\Phi}^{(-H-1)}_{{\boldsymbol \ell},{\boldsymbol s}})^T$}. \\
\end{proposition}

\begin{remark}
As shown from the proof, the simple form \eqref{eq:WavSeries01} holds because the orthonormal basis of $L^2([0,1]^2)$ that we have considered (the periodic wavelet basis) involves a unique scaling function which is the indicator function $\mathbb{I}_{[0,1]^2}(\xx)$. It is therefore clear that the representation \eqref{eq:WavSeries01} remains valid for any other wavelet basis of $L^2([0,1]^2)$ having the indicator function as its unique scaling function. This is in particular the case for the folded wavelet basis \cite{mallat2008wavelet}.\\
\end{remark}

\begin{remark}
Following Remark~\ref{rem1},  it is easy to adapt the proof of Proposition~\ref{corrol01} to the case $H>1$ and deduce that   \eqref{eq:WavSeries01}  remains also valid for $H>1$, since the assumption $\int_{[0,1]^2} \tvt(\xx)d\xx=0$ removes all  constant terms in the wavelet expansion. Moreover, when $H$ is an integer, there exists no clear representation of the divergence-free fBm. In particular this case is excluded from the definition in \cite{Tafti10}. Since the representation  \eqref{eq:WavSeries01}  is well defined for any $H$, we choose to extend  by continuity  \eqref{eq:WavSeries01} to integer values of $H$. Note that this convention makes $\tvt$ satisfy all conditions required in Definition~\ref{def:fbm} when $H$ is an integer. In particular self-similarity is ensured.\\
\end{remark}

\subsection{Divergence-free   wavelet decomposition} \label{sec:divFreeFracRep}

Since $\tvt$ is continuous and divergence-free, then $\tvt \in L^2_{div}([0,1]^2)$, where  
 $L^2_{div}([0,1]^2)$ denotes the space of divergence-free bi-variate fields in  $[0,1]^2$, \ie
$$L^2_{div}([0,1]^2) \triangleq \{ \ww \in~(L^2([0,1]^2))^2: \textrm{div}\, \ww =0 \}.$$

In this section we  {\color{black} decompose}   $\tvt$ onto a biorthogonal wavelets basis of $L^2_{div}([0,1]^2)$. Such basis is constructed from an orthonormal basis of $L^2([0,1]^2)$, as described below. 

Let us start from the periodized  wavelets $\psi^{per}_{\ell, s}$, $\ell, s \in \mathbb{Z}$ defined in \eqref{psiper}.
The primal divergence-free wavelets $\boldsymbol \Psi_{{\boldsymbol \ell},{\boldsymbol s}} \triangleq (\Psi^1_{{\boldsymbol \ell},{\boldsymbol s}}, \Psi^2_{{\boldsymbol \ell},{\boldsymbol s}})^T$ of the biorthogonal wavelets basis of $L^2_{div}([0,1]^2)$ are  defined  for $({\boldsymbol \ell},{\boldsymbol s})   \in \Omega$ by
\begin{itemize}
\item[-] for  $ {0\! \le \! \ell_1\!<\!2^{s_1-1}, 0\le \!\ell_2\!<\!2^{s_2-1},  s_1,s_2 > 0}:$
 $$\Psi^1_{{\boldsymbol \ell},{\boldsymbol s}}(\xx) =   \psi^{per}_{\ell_1,s_1}(x_1)\psi'^{per}_{\ell_2,s_2}(x_2)\,\,\, \textrm{and}  \,\,
\, \Psi^2_{{\boldsymbol \ell},{\boldsymbol s}}(\xx)=-  \psi'^{per}_{\ell_1,s_1}(x_1)\psi^{per}_{\ell_2,s_2}(x_2), \,\,\,  $$ 
\item[-] for $0\!\le \!\ell_1\!<\!2^{s_1-1},  s_1> 0,\, \ell_2=0, s_2=0:$
 $$\Psi^1_{{\boldsymbol \ell},{\boldsymbol s}}(\xx) =  0\,\,\, \textrm{and}  \,\,
\, \Psi^2_{{\boldsymbol \ell},{\boldsymbol s}}(\xx)=-  \psi'^{per}_{\ell_1,s_1}(x_1)  \mathbb{I}_{[0,1]}(x_2), \,\,\,  $$ 
\item[-] for $\ell_1=0, s_1=0,\, 0\!\le \!\ell_2\!<\!2^{s_2-1},  s_2> 0:$
 $$\Psi^1_{{\boldsymbol \ell},{\boldsymbol s}}(\xx) =    \mathbb{I}_{[0,1]}(x_1)\psi'^{per}_{\ell_2,s_2}(x_2)\,\,\, \textrm{and}  \,\,
\, \Psi^2_{{\boldsymbol \ell},{\boldsymbol s}}(\xx)=0, \,\,\,  $$ 
\end{itemize}
where $\psi'$  denotes the derivative  of $\psi$.  To complete the primal wavelets family, the function  $\boldsymbol \Psi_0(\xx)\triangleq ( \mathbb{I}_{[0,1]^2}(\xx), \mathbb{I}_{[0,1]^2}(\xx))^T$ is superimposed, leading to the primal divergence-free wavelets family $\{\boldsymbol \Psi_{{\boldsymbol \ell},{\boldsymbol s}};{({\boldsymbol \ell},{\boldsymbol s})}   \in \Omega\cup 0 \}$. Note that we have $\boldsymbol\Psi_{{\boldsymbol \ell},{\boldsymbol s}}={\bf curl} [{\Phi}_{{\boldsymbol \ell},{\boldsymbol s}}]$   where $\textrm{{\bf curl}}\triangleq  (\frac{\partial }{\partial x_2},-\frac{\partial }{\partial x_1})^t$ and $\Phi_{{\boldsymbol \ell},{\boldsymbol s}}$ is the orthonormal basis constructed in section~\ref{sec:wave}.

The dual wavelets $\tilde{\boldsymbol\Psi}_{{\boldsymbol \ell},{\boldsymbol s}}\triangleq (\tilde \Psi^1_{{\boldsymbol \ell},{\boldsymbol s}},\tilde \Psi^2_{{\boldsymbol \ell},{\boldsymbol s}})^T$ are then constructed in order to be  biorthogonal to  $\boldsymbol\Psi_{{\boldsymbol \ell},{\boldsymbol s}} $, \ie  $\langle \boldsymbol\Psi_{{\boldsymbol \ell},{\boldsymbol s}}/\tilde{\boldsymbol \Psi}_{{\boldsymbol \ell}',{\boldsymbol s}'}\rangle =\delta_{{\boldsymbol \ell},{\boldsymbol \ell}'}\delta_{{\boldsymbol s},{\boldsymbol s}'}$ for all ${{\boldsymbol \ell},{\boldsymbol \ell}'} ,{\boldsymbol s},{\boldsymbol s}' $. They are given by   $\tilde{\boldsymbol\Psi}_0={\boldsymbol\Psi}_0$ and for $({\boldsymbol \ell},{\boldsymbol s})\in\Omega$, 
\begin{itemize}
\item[-] for  $ {0\! \le \! \ell_1\!<\!2^{s_1-1}, 0\le \!\ell_2\!<\!2^{s_2-1},  s_1,s_2 > 0}:$
$$\tilde \Psi^1_{{\boldsymbol \ell},{\boldsymbol s}}(\xx)=  - \psi^{per}_{\ell_1,s_1}(x_1)\int_{0}^{x_2}\psi^{per}_{\ell_2,s_2}(x) dx \,\,\, \textrm{and}  \,\,
\,  \tilde\Psi^2_{{\boldsymbol \ell},{\boldsymbol s}}(\xx)=  \psi^{per}_{\ell_2,s_2}(x_2)\int_{0}^{ x_1}\psi^{per}_{\ell_1,s_1}(x)dx,$$
\item[-] for $0\!\le \!\ell_1\!<\!2^{s_1-1},  s_1> 0,\, \ell_2=0, s_2=0:$
 $$\tilde\Psi^1_{{\boldsymbol \ell},{\boldsymbol s}}(\xx) =  0\,\,\, \textrm{and}  \,\,
\, \tilde\Psi^2_{{\boldsymbol \ell},{\boldsymbol s}}(\xx)= \mathbb{I}_{[0,1]}(x_2)\int_{0}^{x_1}\psi^{per}_{\ell_1,s_1}(x)dx, $$ 
\item[-] for $\ell_1=0, s_1=0,\, 0\!\le \!\ell_2\!<\!2^{s_2-1},  s_2> 0:$
 $$\tilde\Psi^1_{{\boldsymbol \ell},{\boldsymbol s}}(\xx) =  -\mathbb{I}_{[0,1]}(x_1) \int_{0}^{x_2}\psi^{per}_{\ell_2,s_2}(x)dx\,\,\, \textrm{and}  \,\,
\, \tilde\Psi^2_{{\boldsymbol \ell},{\boldsymbol s}}(\xx)=0.$$ 
\end{itemize}

{\color{black}
The expansion  of $\tvt$ in $L^2_{div}([0,1]^2)$ with respect to the divergence-free biorthogonal wavelet basis described above writes
\begin{align*}
\tvt(\xx)=\sum_{({\boldsymbol \ell},{\boldsymbol s })\in \Omega\cup 0} d_{{\boldsymbol \ell},{\boldsymbol s}}\boldsymbol \Psi_{{\boldsymbol \ell},{\boldsymbol s}}(\xx),
\end{align*}
where $d_{{\boldsymbol \ell},{\boldsymbol s}}\triangleq\langle\tvt/\tilde{\boldsymbol\Psi}_{{\boldsymbol \ell},{\boldsymbol s}}\rangle$ are  the random divergence-free wavelet coefficients. Unlike decomposition \eqref{eq:WavSeries01}, these wavelets coefficients are in general correlated. The following proposition describes their covariance structure under the assumption $\int_{[0,1]^2} \tvt(\xx)d\xx=0$.  Note that in this case,  since $\langle\tvt/\tilde{\boldsymbol\Psi}_0\rangle=0$, the decomposition of $\tvt$ becomes
\begin{align}\label{eq:DivFreeDec}
\tvt(\xx)=\sum_{({\boldsymbol \ell},{\boldsymbol s })\in \Omega} d_{{\boldsymbol \ell},{\boldsymbol s}}\boldsymbol \Psi_{{\boldsymbol \ell},{\boldsymbol s}}(\xx).
\end{align}

\begin{proposition} \label{prop:covDls}
Let $\tvt$ be an  isotropic divergence-free fBm with parameter $H \in (0,1)$. Assume that $\int_{[0,1]^2} \tvt(\xx)d\xx=0$. Then, the  coefficients $d_{{\boldsymbol \ell},{\bf s}}$, $({\boldsymbol \ell}, {\bf s}) \in \Omega$ are zero-mean correlated Gaussian random variables, characterized by the following covariance function:  for any $({\boldsymbol \ell}, {\bf s}),({\boldsymbol \ell}', {\bf s}') \in\Omega$, 
\begin{equation}\label{eq:crosscorrelDls} 
{\bf \Sigma}({{\boldsymbol \ell},{\bf s}},{{\boldsymbol \ell}',{\bf s}'})\triangleq\mathbb{E}[d_{{\boldsymbol \ell},{\bf s}}d_{{\boldsymbol \ell}',{\bf s}'}]=4 (2\pi\sigma)^2 \langle  {\Phi}^{(-H-2)}_{{\boldsymbol \ell},{\bf s}} , {\Phi}^{(-H-2)}_{{\boldsymbol \ell'},{\bf s'}} \rangle,
\end{equation}
where $ {\Phi}^{(-H-2)}_{{\boldsymbol \ell},{\bf s}}$ is defined in \eqref{eq:fracdiff}.\\
\end{proposition}

In the paper remainder, we will assume  $\int_{[0,1]^2} \tvt(\xx)d\xx=0$  to fit the assumptions of Propositions~\ref{corrol01} and  \ref{prop:covDls}.

}




\section{MAP estimation}\label{sec:Bayes}


At this point, we have gathered all  ingredients necessary to formalize properly the problem of estimating the incompressible velocity field $\tvt$ of the introduction by a MAP approach. In this section, we first justify the choice for the likelihood model $p_{\delta y | \tvtr}$ from physically sound consideration. Then, we express the fBm prior for $\tvt$ in terms of the two wavelet expansions presented before. Finally, we deduce the optimization problems to solve in practice in order to obtain the MAP estimates.


\subsection{Likelihood model} \label{sec:likeli}
Solving  a turbulent motion inverse problem consists in recovering  a deformation field, denoted later on by $\tvtr$,  from the observation of an image pair.  To this aim, the estimation classically relies on a likelihood model  linking the unknown  deformation to the observed image pair $\{y_0, y_1\}$.   We first assume that this image pair is the solution $y (\xx,t) \in  \mathcal{C}^1([0,1]^2 \times \mathbb{R})$ taken at times $t_0$ and $t_1$ (with $t_1>t_0$) of the following transport  equation, so-called in the image processing literature ``optic-flow equation'': 
\begin{equation}\label{eq:OFC}
 \left\{\begin{aligned}
&\frac{\partial y}{\partial t}(\xx,t)  + \wwr(\xx,t) \cdot\nabla_{\xx}  y(\xx,t) =0\\
&y(\xx,0)=y_0(\xx)
\end{aligned}\right.,
\end{equation}
where $\wwr(\xx,t)$ is the transportation field that verifies, at any time $t\in\mathbb R$, $\wwr(.,t)\in L^2_{div}([0,1]^2)$ due to the incompressibility constraint.
It is well known  \cite{raviart1983introduction} that we can write $y_0(\xx) =y_1({\bf X}^{t_1}_{t_0}(\xx)) $ where the function $t \rightarrow {\bf X}_{ t_0}^t(\xx)$, known as the characteristic curves of the partial differential equation \eqref{eq:OFC}, is  the solution of the system:
 \begin{equation*}
 \left\{\begin{aligned}
&\frac{d }{d t}{{\bf X}}_{ t_0}^t(\xx)  = \wwr({{\bf X}}_{ t_0}^t(\xx),t) \\
&{{\bf X}}_{ t_0}^{t_0}(\xx)=\xx
\end{aligned}\right..
\end{equation*}
Let us   remark  that system \eqref{eq:OFC}  constitutes  a relevant model  frequently  encountered  in physics, in particular  in fluid mechanics: it describes the non-diffusive advection of a passive scalar by the flow $\wwr$ \cite{Liu08}.   
 Assuming that  $\int_{t_0}^{t_1}\wwr({{\bf X}}_{ t_0}^s(\xx),s)ds=\wwr(\xx,t_0)$ when $dt\triangleq t_1-t_0$ is a small increment, then denoting $ \tvtr(\xx)\triangleq dt\, \wwr(\xx,t_0)$ we  obtain by time integration the  so-called  Displaced Frame Difference (DFD) constraint: $$y_0(\xx) =y_1(\xx+ \tvtr(\xx)).$$ 
 
Moreover, in the context of a  physical experiment,  the observed image pair is likely to be corrupted by  noise measurement errors. 
In this context, a standard  approach  to  estimate $\tvtr$  is to  adopt a probabilistic  framework. 
 We make  the assumption that the quantity  
$y_{1}(\xx+\tvtr(\xx))-y_0(\xx) $ is corrupted by a centered Gaussian noise. More precisely, we assume that the so-called data-term DFD functional
\begin{equation}\label{objFunction}
\delta y( \tvtr)\triangleq\frac{1}{2}\|y_{1}(\xx+\tvtr(\xx))-y_0(\xx)\|^2; \hspace{0.25cm} \tvtr \in L_{div}^2([0,1]^2),\; y_0, y_1  \in L^2([0,1]^2),
 \end{equation} 
follows an exponential law, so that the {likelihood} model writes 
 \begin{equation}\label{eq:likli_}
p_{\delta y | \tvtr} = \beta\exp^{-\beta \delta y(\tvt)}, 
\end{equation} 
where $\beta$ is a positive constant.


A straightforward criterion to estimate deformation field $\tvtr$ from the observation of $\{y_0, y_1\}$ is to maximize the likelihood: ${\arg \min}_{\tvtr  \in L^2_{div}}\,\delta y( \tvtr)$. Although  searching the deformation field $\tvtr$ in $L^2_{div}$ instead of $(L^2)^2$ reduces by a factor two the number of degrees of freedom,  this  problem remains  ill-conditioned. This is due to the so-called aperture problem \cite{KadriHarouna12,Yuan07}. Therefore, a Bayesian framework introducing prior information for $\tvtr$ is needed for regularization of the solution. \\

\subsection{Prior models}\label{sec:prior}
 As explained in introduction, we choose as a prior for $\tvt$ the divergence-free fBm described in section~\ref{app:spectralDef}. 
 From the wavelets decomposition \eqref{eq:WavSeries01} or {\color{black} \eqref{eq:DivFreeDec},  this prior can}  
\begin{itemize}
\item[(i)]  either be represented by a Leray projection  of a fractional Laplacian wavelet series, whose coefficients are  distributed according to independent  standard normal distributions. 
\item[(ii)]  or by divergence-free wavelet series, whose coefficients are correlated according to \eqref{eq:crosscorrelDls}.
\end{itemize}
In practice, the images $y_0$ and $y_1$ are of size $n \times n$ pixels. Accordingly we will truncate the series \eqref{eq:WavSeries01} and {\color{black} \eqref{eq:DivFreeDec}}  with respect to the index set $$\Omega_n \triangleq \{ {{\boldsymbol \ell},{\boldsymbol s}} \in   \Omega; s_1,s_2 \le s_n\triangleq log_2(n) \}.$$ In other words,  wavelet coefficients revealing scales smaller than the pixel size will be neglected. 
 We thus consider the  vector  of  fractional Laplacian wavelet coefficients ${\boldsymbol \epsilon_n}=({ \epsilon}_n^1, { \epsilon}_n^2)^T$ with ${ \epsilon}_n^i=\{\epsilon^i_ {{\boldsymbol \ell},{\boldsymbol s}};\,({{\boldsymbol \ell},{\boldsymbol s}})\in \Omega_n\}$ and the vector of divergence-free wavelet coefficients  $ d_n=\{d_ {{\boldsymbol \ell},{\boldsymbol s}};\,({{\boldsymbol \ell},{\boldsymbol s}})\in \Omega_n\}$. 
 
 {\color{black}
 Further, we assume in practice periodic boundaries, so that the wavelets considered so far, defined on $[0,1]^2$, are extended by periodization to $\mathbb R^2$. Accordingly, the Fourier transform applied so far becomes the sequence of Fourier coefficients, and the inverse Fourier transform a Fourier series. They will  actually be implemented in the next section by a simple Fast Fourier Transform. 
 
 In order to express the prior for $\tvt$ in this finite dimensional and periodic  setting, we need to introduce some notation.
 For any  function $f$ in $L^2[0,1]^2$,  considering its extension by periodization to $\mathbb R^2$, we denote its Fourier coefficients by
 $$\mathcal F_{per}(f)(\kk) = \int_{{[0,1]}^2}  f(\xx) e^{-2\pi i\kk \cdot \xx} d\xx,\quad  \forall \kk \in \mathbb{Z}^2,$$
and for any set of coefficients  $\{\hat f (\kk)\}_{ \kk \in \mathbb{Z}^2}$ in $\ell^2( \mathbb{Z}^2)$, we denote  the Fourier series reconstruction operator on the square  by  
$$\mathcal F_{per}^{-1}(\{\hat f (\kk)\}_{ \kk \in \mathbb{Z}^2} )(\xx)= \sum_{\kk \in \mathbb{Z}^2} e^{2\pi i{\bf k} \cdot \xx}\hat f(\kk),\quad  \forall \xx \in [0,1]^2.$$
For a  bivariate function  $\ww=(\ww_1,\ww_2)^T$, $\mathcal F_{per} (\ww) \triangleq (\mathcal F_{per}(\ww_1),\mathcal F_{per}(\ww_2))^T$ and similarly for $\mathcal F_{per}^{-1}$. The Leray projector in this periodic setting is denoted by $\mathcal{P}_{per}$. It maps  bi-variate functions $\ww$ in $(L^2[0,1]^2)^2$ onto $L_{div}^2([0,1]^2)$:
 \begin{align}\label{eq:ProjLerayDefPer}
 \mathcal{P}_{per} {\ww} \triangleq \mathcal F_{per}^{-1}\left( \left\{\left[ \mathbf{I} - \frac{\kk\kk^T}{ \| {\kk }\|^2}\right] \mathcal F_{per} (\ww)(\kk)\right\}_{\kk\in\mathbb Z^2}\right).
 \end{align}

 For any $H>0$ we introduce operator
${\underline{\boldsymbol \Phi}^{(-H-1)}_{  n}}$ from $(\ell^2( \Omega_n))^2$ into $(L^2([0,1]^2))^2$  defined for any $  {\boldsymbol \epsilon}_n \in ( \ell^2( \Omega_n))^2$ by
\begin{align}\label{eq:operatorFrac}
{\underline{\boldsymbol \Phi}^{(-H-1)}_{  n}}{\boldsymbol \epsilon}_n  \triangleq\sum_{({\boldsymbol \ell},{\boldsymbol s })\in  \Omega_n}  \mathcal F_{per}^{-1} \left(\binom{\epsilon^1_{{\boldsymbol \ell},{\boldsymbol s}}  \| {\bf k } \| ^{-H-1}\mathcal F_{per} ( {\Phi}_{{\boldsymbol \ell},{\bf s}})(\kk)} {\epsilon^2_{{\boldsymbol \ell},{\boldsymbol s}}  \| {\bf k } \| ^{-H-1}\mathcal F_{per} ( {\Phi}_{{\boldsymbol \ell},{\bf s}})(\kk)}_{\kk\in\mathbb Z^2}\right).
\end{align}

We introduce analogously  operator 
${\underline{\boldsymbol \Psi}}_n:  \ell^2( \Omega_n)\rightarrow L^2_{div}([0,1]^2)$ defined for any $d_n \in \ell^2( \Omega_n)$ by 
\begin{align}\label{eq:operatorDiv0}
{\underline{\boldsymbol \Psi}}_n d_n = \sum_{({\boldsymbol \ell},{\boldsymbol s })\in  \Omega_n} d_{{\boldsymbol \ell},{\boldsymbol s }} \boldsymbol\Psi_{{\boldsymbol \ell},{\boldsymbol s}},
\end{align}
where the vector  $d_n=\{d_{{\boldsymbol \ell},{\boldsymbol s }}$, $({\boldsymbol \ell},{\boldsymbol s })\in  \Omega_n\}$ is a  zero-mean Gaussian vector with covariance given by an adaptation of \eqref{eq:crosscorrelDls} to the periodic framework. 
 Specifically its  covariance matrix is denoted by ${\bf \Sigma}_n$  whose element at row $({\boldsymbol \ell},{\boldsymbol s})\in  \Omega_n$ and column $({\boldsymbol \ell}',{\boldsymbol s}')\in  \Omega_n$ is
\begin{align}\label{defsigma}
{\Sigma}&({{\boldsymbol \ell},{\boldsymbol s}},{\boldsymbol \ell}',{\boldsymbol s}')=\nonumber\\
&\textcolor{black}{\sigma^2} \langle \left\{ \| {\bf k } \| ^{-H-2}\mathcal F_{per} ( {\Phi}_{{\boldsymbol \ell},{\bf s}})(\kk)\right\}_{\kk\in\mathbb Z^2} , \left\{ \| {\bf k } \| ^{-H-2}\mathcal F_{per} ( {\Phi}_{{\boldsymbol \ell'},{\bf s'}})(\kk)\right\}_{\kk\in\mathbb Z^2}\rangle.
\end{align}
where $\langle.,.\rangle$ denotes here the scalar product in $\ell^2(\mathbb Z^2)$. 
Finally, to compute the probability density of  $d_n$, we need to invert  ${\bf \Sigma}_n$. We approach this inverse matrix by ${\bf \Sigma}^{-1}_n$ composed of elements
\begin{align}\label{definvsigma}
{ \Sigma}^{-1}&({{\boldsymbol \ell},{\boldsymbol s}},{\boldsymbol \ell}',{\boldsymbol s}')=\nonumber \\
&\textcolor{black}{\sigma^{-2}}  \langle \left\{ \| {\bf k } \| ^{H+2}\mathcal F_{per} ( {\Phi}_{{\boldsymbol \ell},{\bf s}})(\kk)\right\}_{\kk\in\mathbb Z^2} , \left\{ \| {\bf k } \| ^{H+2}\mathcal F_{per} ( {\Phi}_{{\boldsymbol \ell'},{\bf s'}})(\kk)\right\}_{\kk\in\mathbb Z^2}\rangle.
\end{align}
This choice is justified by the following lemma, proving that this approximation becomes accurate for $n$ sufficiently large.
\begin{lemma}\label{prop:InvCovDls}
When $n\to\infty$, the matrices ${\bf\Sigma}_n$ and ${\bf\Sigma}_n^{-1}$ become  operators from $ \ell^2(\Omega)$ to $ \ell^2(\Omega)$, that are inverses of each other.  \\
 \end{lemma}

 Henceforth, we use  the two following priors: 
 \begin{itemize}
\item[(i)] based on fractional Laplacian wavelet decomposition \eqref{eq:WavSeries01}:
 \begin{align}\label{priorMod}
\tvt_n=\mathcal P_{per} \left({\underline{\boldsymbol \Phi}^{(-H-1)}_{  n}} {\boldsymbol \epsilon_n}\right),\end{align}
with $p_{\boldsymbol \epsilon_n} = \frac{1}{(2\pi\sigma^2)^{n/2}} e^{-\frac{1}{2\sigma^2} {\boldsymbol \epsilon_n^T}  {\boldsymbol \epsilon_n}}$ ;\\
\item[(ii)] based on divergence-free decomposition   {\color{black} \eqref{eq:DivFreeDec}}:
\begin{align}\label{priorMod2}
\tvt_n={\underline{\boldsymbol \Psi}}_n  d_n,
\hspace{0.5cm} \textrm{with}\,\,\,\,\,\, p_{  d_n} = \frac{1}{(2\pi)^\frac{n}{2} \det^{\frac{1}{2}}({\bf \Sigma}_n)} e^{-\frac{1}{2} {  d_n}^T {\bf \Sigma}_n^{-1}  { d_n}} .
\end{align}
\end{itemize}
These two priors are adaptations  to the finite-dimensional and periodic setting of the decompositions   \eqref{eq:WavSeries01} and  \eqref{eq:DivFreeDec}. For the sake of conciseness, some multiplicative constants have been removed to be included in $\sigma^2$, which becomes a tuning parameter in the MAP procedure described below.

 }

\subsection{Maximum a posteriori estimation}\label{sec:post}

      From \eqref{priorMod} or \eqref{priorMod2}, 
 $\tvt_n$ reduces to the knowledge of wavelet coefficients ${\boldsymbol \epsilon_n}$ or $ d_n$. Let us rewrite the likelihood in terms of these  coefficients:
\begin{equation*}
p_{\delta y | {\boldsymbol \epsilon}_n} =p_{\delta y | {\boldsymbol  d_n}} = \beta\exp^{-\beta \delta y}, 
\end{equation*} 
with the DFD data term  \eqref{objFunction} rewritten as: 
\begin{align}\label{JL}\delta y(\cdot)=\frac{1}{2}\|\bar y_1(\xx,\cdot) -y_0(\xx)\|^2,\end{align}
where
{\color{black}
\begin{align}\label{Imrecal}
\bar y_1(\xx,{\boldsymbol \epsilon_n})\triangleq y_1\left(\xx+\mathcal P_{per} \left({\underline{\boldsymbol \Phi}^{(-H-1)}_{  n}} {\boldsymbol \epsilon_n}\right)(\xx)\right) \hspace{0.3cm}\textrm{or} \hspace{0.3cm} \bar y_1(\xx, d_n)\triangleq y_1\left(\xx+{\underline{\boldsymbol \Psi}}_n  d_n(\xx)\right).
\end{align}
}
The MAP estimates are  defined by:
\begin{align}\label{objFunction5}
{\boldsymbol \epsilon_n}^*=&\underset{ {\boldsymbol \epsilon}_n}{\arg \max}\,p_{\delta y | {\boldsymbol \epsilon}_n} p_{{\boldsymbol \epsilon_n}}\hspace{0.3cm}\textrm{and} \hspace{0.3cm} {  d}_n^*=\underset{  d_n}{\arg \max}\,p_{\delta y | {  d}_n} p_{{  d}_n}.
\end{align}
 Solving the MAP problems  \eqref{objFunction5} is equivalent to minimize 
 minus the logarithm of the posterior distributions:
 
 \begin{itemize}
\item[i)] with respect to fractional Laplacian wavelet coefficients 
\begin{align}
{\boldsymbol \epsilon_n}^*&=\underset{\boldsymbol \epsilon_n}{\arg \min}\{ \delta y({\boldsymbol \epsilon_n}) +\mathcal{R}({\boldsymbol \epsilon_n})\}\label{objFunction6}\\
\mathcal{R}({\boldsymbol \epsilon_n})&\triangleq
{\color{black} \frac{1}{2\beta\sigma^2}} {\boldsymbol \epsilon}_n^T  {\boldsymbol \epsilon}_n \label{eq:regFracLap},
\end{align} 
\item[ii)]with respect to   divergence-free wavelet coefficients \begin{align}
 d_n^*&=\underset{\boldsymbol  d_n}{\arg \min}\{ \delta y({d_n}) +\mathcal{R}({d_n})\}\label{objFunction6_}\\
\mathcal{R}( d_n)&\triangleq 
{\color{black}\frac{1}{2\beta}}  {  d_n}^T {\bf \Sigma}_n^{-1}  { d_n}\label{eq:regDivFree},
\end{align} 
\end{itemize}
where $\mathcal{R}$'s are so-called  regularizers.

In the following we will assume that the Hurst exponent $H$  is known.   Posterior models described above are thus characterized by two free-parameters, namely $\beta$ and $\sigma$. However, MAP estimates  \eqref{objFunction6} and \eqref{objFunction6_} only depend on the product of these two parameters,  forming the so-called regularization parameter $\frac{1}{\beta\sigma^2}$. This regularization parameter  explicitly appears in \eqref{eq:regFracLap} while it is partially hidden in the covariance matrix inverse in \eqref{eq:regDivFree}.  Let us  mention that an empirical study   shows a low sensitivity  to  the choice of the regularization parameter  (see  section~\ref{sec:results}).  Nevertheless, estimation techniques exist when no prior knowledge is available for the adjustment of  Hurst exponent  or regularization parameter \cite{Heas12a,Mackay92,Stein81}.

\section{Optimization}\label{sec:optim}

  In this section, we introduce algorithms to solve   \eqref{objFunction6} and \eqref{objFunction6_}. To deal with these  non-convex optimization problems, we  use a Limited-memory  Broyden-Fletcher-Goldfarb-Shanno~(LBFGS) procedure, \ie  a  quasi-Newton method with a  line search strategy, subject to  the strong Wolf conditions \cite{Nocedal99}.
 Moreover,  as suggested in \cite{Derian12wavelet} we propose to enhance optimization  by solving a sequence of nested   problems, in which  the MAP solution  is sequentially sought within higher resolution spaces.  More precisely, wavelet coefficients are estimated sequentially from the coarsest scale $2^{0}$ to the  finest one $2^{-s_n}$. At each scale,  problems \eqref{objFunction6} and \eqref{objFunction6_} are solved by the LBFGS method with respect to a growing subset of wavelet coefficients. This subset  includes all coefficients from the coarsest scale to the current one $2^{-s}$, coefficients estimated at previous coarser scales  being used as the initialization point of the gradient descent. This strategy enables to update  those coarser coefficients while estimating new details at current scale $2^{-s}$. 
 
To implement the above procedure, the functional gradient and the functional itself in \eqref{objFunction6} and \eqref{objFunction6_} need to be evaluated at any point ${\boldsymbol \epsilon_n}$ and $d_n$. Note that once the functional gradient is determined, the functional value is  simple to deduce:    the evaluation of \eqref{Imrecal} needed by \eqref{JL} will be  a precondition to the computation of the DFD functional gradient  while  \eqref{eq:regFracLap} or \eqref{eq:regDivFree} may be derived from their gradients by simple  scalar product  with the vector of wavelet coefficients ${\boldsymbol \epsilon_n}$ or $d_n$.  

The next section provides  algorithms to compute exactly the functional gradient for  the two different wavelet representations. Besides, an approached gradient computation  is proposed to accelerate the algorithm in the case of the divergence-free wavelet representation.  

\subsection{Projected fractional  wavelet series}\label{sec:optim_frac}

\noindent

\begin{proposition}  \label{lemma:specGrad}
{\color{black}Let $\mathcal{P}^i_{per} $ denote the $i$-th row of the Leray projector $\mathcal{P}_{per}$ defined by \eqref{eq:ProjLerayDefPer}.} The gradient  of functional minimized in \eqref{objFunction6} with respect to $\epsilon^i_{{\boldsymbol \ell},{\boldsymbol s}}$ is $ \partial_{\epsilon^i_{{\boldsymbol \ell},{\boldsymbol s}}} \delta y({\boldsymbol \epsilon_n}) +\partial_{\epsilon^i_{{\boldsymbol \ell},{\boldsymbol s}}} \mathcal{R}({\boldsymbol \epsilon_n})$ where   for  $i=1,2$ and for all $({{\boldsymbol \ell},{\boldsymbol s}}) \in \Omega_n$:
\begin{align}
&\partial_{\epsilon^i_{{\boldsymbol \ell},{\boldsymbol s}}} \delta y({\boldsymbol \epsilon_n}) = \nonumber   \\
&\langle {\color{black} \mathcal{F}_{per}^{-1} \left( \left( \|\kk \|^{\!-\!H\!-\!1\!} \mathcal{F}_{per} \left( \mathcal{P}^i_{per} [(\bar y_1(\cdot,{\boldsymbol \epsilon_n}) -y_0(\cdot) )\nabla    \bar y_1(\cdot,{\boldsymbol \epsilon_n})] \right) (\kk)\right)_{\kk \in \mathbb{Z}^2}\right)}, {{\Phi}}_{{\boldsymbol \ell},{\boldsymbol s}} \rangle \label{gradAdjoint}\\
&\partial_{\epsilon^i_{{\boldsymbol \ell},{\boldsymbol s}}} \mathcal{R}({\boldsymbol \epsilon_n}) ={\color{black}\frac{1}{\beta{\sigma^2}}}{\epsilon^i_{{\boldsymbol \ell},{\boldsymbol s}}}.\label{gradJp}
\end{align}
\end{proposition}

Based on these formula, we derive a spectral method for the computation of the gradient of  functional minimized in \eqref{objFunction6}. It is  based on FFT and   FWT with recursive filter banks:\\
\begin{algo}\label{algo1}
{ {\bf(functional gradient w.r.t ${\boldsymbol \epsilon_n}$)} 
\begin{itemize} 
\item[i)] compute the FFT of the components of $(\bar y_1 -y_0 )\nabla    \bar y_1$,
\item[ii)] {\color{black} apply operator $\|\kk \|^{\!-\!H\!-\!1\!}$ }and  Leray projection in Fourier domain to get\\ {\color{black}$ \left( \|\kk \|^{\!-\!H\!-\!1\!} \mathcal{F}_{per} \left( \mathcal{P}^i_{per} [(\bar y_1(\cdot,{\boldsymbol \epsilon_n}) -y_0(\cdot) )\nabla    \bar y_1(\cdot,{\boldsymbol \epsilon_n})] \right) (\kk)\right)_{\kk \in \mathbb{Z}^2}$, for $i=1,2$}
\item[iii)] compute the inverse FFT 
\item[iv)] decompose each component by FWT using the orthogonal wavelets $\Phi_{{\boldsymbol \ell},{\boldsymbol s}}$  to obtain the  data-term gradient \eqref{gradAdjoint},
\item[v)]  Derive functional gradient  by adding vector \eqref{gradJp}  to the data-term gradient.\\
\end{itemize}
}
\end{algo}

\noindent
In order to evaluate $\bar y_1$ or its gradient in the above algorithm,  one needs to reconstruct the unknown $\tvtr_n $ \eqref{priorMod} appearing   in \eqref{Imrecal} from the fractional Laplacian wavelet coefficients  ${\boldsymbol \epsilon_n}$.   This can  be done by the following spectral computation\footnote{Let us remark that this reconstruction essentially differs from a direct spectral fBm generation method which is known to bring aliasing side-effects, see \cite{Bardet03}.} of the inverse fractional wavelet transform. Indeed, by commuting  Leray projector with fractional integration, from \eqref{priorMod}  we get:
\textcolor{black}{
\begin{align}\label{recUFourier}
\tvtr_n= \mathcal F_{per}^{-1} \left(\left( \| {\bf k } \| ^{-H-1}\left[ \mathbf{I} - \frac{\kk\kk^T}{ \| {\kk }\|^2}\right] \mathcal F_{per} ({\underline{\boldsymbol\Phi}^{(0)}_{  n}} {\boldsymbol \epsilon}_n )(\kk)\right)_{\kk\in\mathbb Z^2}\right),\end{align}
}where we recall that  ${\underline{\boldsymbol\Phi}^{(0)}_{  n}}$ is defined in  \eqref{eq:operatorFrac}. In practice  ${\underline{\boldsymbol\Phi}^{(0)}_{  n}} {\boldsymbol \epsilon}_n$ is obtained on a finite grid of points (namely the $n\times n$ pixels), {\color{black}  so that   Fourier  coefficients and  Fourier series in \eqref{recUFourier} are approximated in practice by  FFT and inverse FFT.}
This yields the following reconstruction algorithm:\\

 \begin{algo}{{\bf(reconstruction of fBm from $ {\boldsymbol \epsilon}_n$)}\label{algo2} 
\begin{itemize}
\item[i)]  reconstruct  $ {\underline{\boldsymbol\Phi}^{(0)}_{  n}} {\boldsymbol \epsilon}_n$ from ${\boldsymbol \epsilon_n}$ by  inverse FWT of each component  using  orthogonal wavelets $\{\Phi_{{\boldsymbol \ell},{\boldsymbol s}};({{\boldsymbol \ell},{\boldsymbol s}})\in \Omega_n\}$,
\item[ii)]  compute  FFT of the two components   of the latter function,
\item[iii)]  compute  Leray projection  and fractional differentiation in Fourier domain,
\item[iv)] compute  inverse FFT of the two components to obtain $\tvtr_n$. \\
\end{itemize}
}\end{algo}

Algorithms \ref{algo1} and \ref{algo2} yield the ingredients necessary to approach the MAP estimate ${\boldsymbol \epsilon^*_n}$ with a gradient descent method of theoretical complexity bounded by the FFT algorithm in $\mathcal{O}(n\log n)$. Nevertheless,  the computation bottleneck comes mainly from the number of transforms  that are required at  each gradient decent step: 4 FFT,  4 inverse FFT, 2 FWT and 2 inverse FWT.

\subsection{Divergence-free wavelet series}

\subsubsection{Exact method}\label{sec:SecDiv0Optim}

For notational convenience we define  operators ${\underline{\boldsymbol \Phi}^{i,(-H-1)}_{  n}}: \ell^2( \Omega_n)\rightarrow L^2([0,1]^2)$ for $i=1,2$ as the two components of the  operator defined in \eqref{eq:operatorFrac}:    
\begin{align}\label{eq:operatorFracNewDef}
\begin{pmatrix} {\underline{\boldsymbol \Phi}^{1,(-H-1)}_{  n}} \epsilon^1_n \\{\underline{\boldsymbol \Phi}^{2,(-H-1)}_{  n}}\epsilon^2_n  \end{pmatrix} \triangleq{\underline{\boldsymbol \Phi}^{(-H-1)}_{  n}} {\boldsymbol \epsilon}_n .
\end{align}
Note that we have ${\underline{\boldsymbol \Phi}^{1,(-H-1)}_{  n}}={\underline{\boldsymbol \Phi}^{2,(-H-1)}_{  n}}$.

\begin{proposition}\label{prop:noApproxDiv0Prior}
The gradient  of functional minimized in \eqref{objFunction6_} with respect to $d_{{\boldsymbol \ell},{\boldsymbol s}}$  is $ \partial_{d_{{\boldsymbol \ell},{\boldsymbol s}}} \delta y({ d_n}) +\partial_{d_{{\boldsymbol \ell},{\boldsymbol s}}} \mathcal{R}({ d_n})$ where for all $({{\boldsymbol \ell},{\boldsymbol s}}) \in \Omega_n$:
\begin{align}
\partial_{d_{{\boldsymbol \ell},{\boldsymbol s}}} \delta y({ d_n}) &=\langle  (\bar y_1 -y_0 )\nabla    \bar y_1 /{{\boldsymbol \Psi}}_{{\boldsymbol \ell},{\boldsymbol s}} \rangle\label{gradJlDivFree} 
\end{align}
and 
\begin{align}\label{eq:gradfuncDiv0Noapprox}
 \partial_{d_{{\boldsymbol \ell},{\boldsymbol s}}} \mathcal{R}({ d_n})
 &={\color{black}\frac{1}{\beta\sigma^2}}\langle  {\underline{\boldsymbol \Phi}^{1,{(H+2)}}_{  n}} d_n , {\Phi}^{(H+2)}_{{\boldsymbol \ell},{\boldsymbol s}} \rangle.
 \end{align}
\end{proposition}

Based on the previous result, we derive a spectral method for the computation of the gradient of  functional minimized in \eqref{objFunction6_}. It is  based again on FFT and   FWT with recursive filter banks:\\
\begin{algo}\label{algo5}
{ {\bf(functional gradient w.r.t $ d_n$)} 
\begin{itemize} 
\item[i)] decompose $  (\bar y_1 -y_0 )\nabla    \bar y_1 $ by FWT using  dual divergence-free wavelets $\{\tilde {\boldsymbol\Psi}_{{\boldsymbol \ell},{\boldsymbol s}};({{\boldsymbol \ell},{\boldsymbol s}})\in \Omega_n\}$  to obtain the  data-term gradient \eqref{gradJlDivFree},
\item[ii)] compute  inverse FWT  of ${ d_n}$ using  orthogonal wavelets $\{\Phi_{{\boldsymbol \ell},{\boldsymbol s}};({{\boldsymbol \ell},{\boldsymbol s}})\in \Omega_n\}$.
{\color{black} \item[iii)] compute  FFT and apply the operator\footnote{
This supposes $2H+4$ differentiability of  $\psi$. If we only assume {\color{black} $\max(2H,\,H+2)$} differentiability,  algorithm~\ref{algo5} can be modified as described in Appendix~\ref{app:Algo5Modif}.} $\| \kk\| ^{2H+4}$.
\item[iv)] compute  inverse FFT  to get  $ \mathcal{F}^{-1}_{per}\left(\left( \| \kk\| ^{2H+4}\mathcal{F}_{per}({\underline{\boldsymbol \Phi}_{  n}^{1,(0)}} d_n )\right)_{\kk \in \mathbb{Z}^2}\right)$.}
\item[v)] compute  FWT using  orthogonal wavelets $\{\Phi_{{\boldsymbol \ell},{\boldsymbol s}};({{\boldsymbol \ell},{\boldsymbol s}})\in \Omega_n\}$  and get  \eqref{eq:gradfuncDiv0Noapprox},
\item[vi)]  derive functional gradient  by adding   \eqref{eq:gradfuncDiv0Noapprox} to the data-term gradient.\\
\end{itemize}
}
\end{algo}

\noindent

    Reconstruction of $\tvt_n$ from coefficients  $ d_n$ is  needed to compute $\bar y_1$. It is easily done by an inverse FWT. \\
 \begin{algo}{{\bf(reconstruction of fBm from $  d_n$)}\label{algo4} 
\begin{itemize}
\item[i)]  Reconstruct  $\tvtr_n= {\underline{\boldsymbol\Psi}}_n  d_n$ from $ d_n$ by  inverse FWT using   divergence-free wavelets $\{ {\boldsymbol\Psi}_{{\boldsymbol \ell},{\boldsymbol s}};({{\boldsymbol \ell},{\boldsymbol s}})\in \Omega_n\}$.\\
\end{itemize}
}\end{algo}

Algorithms \ref{algo5} and \ref{algo4} yield the ingredients necessary to approach the MAP estimate ${ d^*_n}$ with a gradient descent method of theoretical complexity bounded again by the FFT algorithm in $\mathcal{O}(n\log n)$. However,   it is less  time-consuming than the approach based on fractional Laplacian wavelet series since it requires for each gradient decent step: 1 FFT,  1 inverse FFT, 2 FWT and 2 inverse FWT. Moreover let us remark that a simple FWT with recursive filter banks corresponding to the dual divergence-free wavelet basis \cite{Lemariet92} is required to compute the data-term gradient.  

\subsubsection{Approach method}\label{sec:SpatialOptim}

In this section, we approach  the divergence-free  regularizer gradient \eqref{eq:gradfuncDiv0Noapprox} in terms of  matrix products, which will make computation of the gradient more efficient since we avoid intensive use of FFT and FWT as in algorithm~\ref{algo5}.  
From Proposition~\ref{prop:noApproxDiv0Prior} and the Parseval formula, we have for all $({\boldsymbol \ell},{\boldsymbol s} )\in  \Omega_n:$
{\color{black}
  \begin{align}\label{eq:conec1}
&{\beta\sigma^2} \partial_{d_{{\boldsymbol \ell},{\boldsymbol s}}} \mathcal{R}({ d_n})  \\
&=\sum_{({\boldsymbol \ell}',{\boldsymbol s}' )\in  \Omega_n} d_{{\boldsymbol \ell}',{\boldsymbol s}'}\langle 
\left( {\|\kk\|}^{H+2} \mathcal F_{per} ( {\Phi}_{{\boldsymbol \ell}',{\boldsymbol s}'})(\kk)\right)_{\kk\in\mathbb Z^2},  
\left( {\|\kk\|}^{H+2} \mathcal F_{per} ( {\Phi}_{{\boldsymbol \ell},{\boldsymbol s}})(\kk)\right)_{\kk\in\mathbb Z^2}
 \rangle. \nonumber
\end{align}}
We hereafter derive a separable approximation of \eqref{eq:conec1}
in terms of  one-dimensional connection coefficients  $f_ {{ \ell'},{ s'},{ \ell},{ s}} ^{(\alpha)}$ defined  for all $({{ \ell'},{ s'}),({ \ell},{ s}}) \in \Omega_n$ as 
 \begin{equation}\label{eq:defCoeffConnec}
   f_ {{ \ell'},{ s'},{ \ell},{ s}} ^{(\alpha)}=
 \left\{\begin{aligned}
&{\color{black}(2\pi)^{-2\alpha}}\langle \psi^{per}_{{ \ell'},{ s'}}, {\left(\frac{-\partial^2}{\partial x^2}\right)}^{\alpha}{\psi}^{per}_{{ \ell},{ s}} \rangle \hspace{0.3cm} \textrm{for}\hspace{0.3cm}0< { s}, s'  \le s_n ,\, 0\le\ell \!<\!2^{s-1}, 0\le \ell'\!<\!2^{s'-1}\\
&{\color{black}(2\pi)^{-2\alpha}}\langle \psi^{per}_{{ \ell'},{ s'}}, {\left(\frac{-\partial^2}{\partial x^2}\right)}^{\alpha}{\mathbb{I}_{[0,1]}} \rangle \hspace{0.3cm} \textrm{for}\hspace{0.3cm}{ s=0},  0< s'  \le s_n ,\,  l=0, 0\le  \ell'\!<\!2^{s'-1}\\
&{\color{black}(2\pi)^{-2\alpha}}\langle \mathbb{I}_{[0,1]}, {\left(\frac{-\partial^2}{\partial x^2}\right)}^{\alpha}{\psi}^{per}_{{ \ell},{ s}} \rangle \hspace{0.3cm} \textrm{for}\hspace{0.3cm}0< s  \le s_n, s'=0,\, 0\le  \ell\!<\!2^{s-1},  { l'=0}\\
&{\color{black}(2\pi)^{-2\alpha}}\langle \mathbb{I}_{[0,1]}, {\left(\frac{-\partial^2}{\partial x^2}\right)}^{\alpha}\mathbb{I}_{[0,1]} \rangle \hspace{0.3cm} \textrm{for}\hspace{0.3cm}{ s=0,s'=0,l=0,l'=0}\\
\end{aligned}\right..
\end{equation}
Note that  for any fixed $\alpha\leq0$ (resp. $\alpha \geq 0$), $f_ {{ \ell'},{ s'},{ \ell},{ s}} ^{(\alpha)}$  exists whenever $\psi$ possesses sufficient vanishing moments (resp. is sufficiently differentiable).
{\color{black}The two-dimensional scalar products of $\ell^2(\mathbb{Z}^2)$ in \eqref{eq:conec1} can be written $\langle 
 \mathcal F_{per} ( {\Phi}_{{\boldsymbol \ell}',{\boldsymbol s}'}),   {\|\kk\|}^{2(H+2)}
 \mathcal F_{per} ( {\Phi}_{{\boldsymbol \ell},{\boldsymbol s}})
 \rangle$.}
If $H \in \mathbb{N}$,    Newton's binomial theorem applies: 
\begin{align} \label{eq:splitSeparable0}
 {\|\kk\|}^{2(H+2)}=\frac{1}{2}\sum_{i=0}^{ H+2  }\binom{H+2}{i}\left( k_1^{2(H+2-i)} k_2 ^{2i}+ k_1^{2i} k_2^{2(H+2-i)}\right), 
\end{align} 
where $$\binom{H+2}{i} \triangleq (H+2)(H+2-1) ... (H+2-i +1)/i! \; .$$
So if  $H \in \mathbb{N}$, plugging \eqref{eq:splitSeparable0} into \eqref{eq:conec1} shows that $\partial_{d_{{\boldsymbol \ell},{\boldsymbol s}}}\mathcal{R}({ d_n})$ can be expressed in a separable form: 
\begin{align}\label{Rseparable}
\partial_{d_{{\boldsymbol \ell},{\boldsymbol s}}} \mathcal{R}({ d_n})=
& {\color{black}\frac{1}{2\beta\sigma^2}} \sum_{i=0}^{H+2}\binom{H+2}{i} \bigg( \underset {{ \ell_1',s_1'}}{\sum} f_ {{ \ell_1'},{ s_1'},{ \ell_1},{ s_1}} ^{(H+2-i)}\underset {{ \ell_2',s_2'}}{\sum}d_ {{ \ell_1',s_1'},{ \ell_2',s_2' ,}}   f_ {{ \ell_2'},{ s_2'},{ \ell_2},{ s_2}} ^{(i)} 
\nonumber\\
&\hspace{3cm}+ \underset {{ \ell_1',s_1'}}{\sum} f_ {{ \ell_1'},{ s_1'},{ \ell_1},{ s_1}} ^{(i)}\underset {{ \ell_2',s_2'}}{\sum}d_ {{ \ell_1',s_1'},{ \ell_2',s_2' ,}}   f_ {{ \ell_2'},{ s_2'},{ \ell_2},{ s_2}} ^{(H+2-i)} \bigg).
\end{align}

The latter formula can be expressed in terms of matrix products. Let $\FF^ {(\alpha)}$ be the matrix of size $n \times n$ composed  at  row index $({ \ell'},{ s'})$ and  column index $({ \ell},{ s})$ by the element $f^{(\alpha)}_ {{ \ell'},{ s'},{ \ell},{ s}}$. Let $\llbracket { d_n}\rrbracket$ the $n \times n$ matrix  whose element at  row index $({{ \ell}_1,{ s}_1})$ and column index $({{ \ell}_2,{ s}_2} ) $ is  $d_{{\boldsymbol \ell},{\boldsymbol s}}$. Equation  \eqref{Rseparable} can then be written
\begin{align}\label{eq:lowDimMatProdGrad0}
\partial_{d_{{\boldsymbol \ell},{\boldsymbol s}}} \mathcal{R}({ d_n}) = & {\color{black}\frac{1}{2\beta\sigma^2}} \left[ \sum_{i=0}^{ H+2}\binom{H+2}{i} \left(\FF^{(H+2-i)^T}\llbracket d_n\rrbracket \FF^ {(i)} +\FF^{(i)^T}\llbracket d_n\rrbracket \FF^ {(H+2-i)}  \right) \right]_{{\boldsymbol \ell},{\boldsymbol s}},
\end{align}
where $[{\mathbf M}]_{{\boldsymbol \ell},{\boldsymbol s}}$ denotes the $({{\boldsymbol \ell},{\boldsymbol s}})$-th element  of matrix ${\mathbf M}$.

In the general case where $H\notin \mathbb{N}$, \eqref{eq:lowDimMatProdGrad0} does not hold anymore, but in the same spirit we consider the following natural approximation\footnote{This approximation may be justified by considering a truncated version of the Newton's generalized formula. However, a control of the approximation error of \eqref{eq:lowDimMatProdGrad} is a technical issue, which is out of the scope of this paper.}:  
\begin{align}\label{eq:lowDimMatProdGrad}
\partial_{d_{{\boldsymbol \ell},{\boldsymbol s}}} \mathcal{R}({ d_n}) \approx &{\color{black}\frac{1}{2\beta\sigma^2}}\left[ \sum_{i=0}^{\lfloor H+2\rfloor}\binom{H+2}{i} \left(\FF^{(H+2-i)^T}\llbracket d_n\rrbracket \FF^ {(i)} +\FF^{(i)^T}\llbracket d_n\rrbracket \FF^ {(H+2-i)}  \right) \right]_{{\boldsymbol \ell},{\boldsymbol s}},
\end{align}
where $\lfloor H+2 \rfloor$ denotes the integer part of $H+2$.

We approximate matrices $\FF^ {(\alpha)}$ by an off-line  FWT of {\it scaling function} connection coefficients, as explained in Appendix~\ref{app1} and summarized in the following algorithm. These scaling function connection coefficients are in turn easily computable as a solution of a linear system, by  adapting Beylkin's fast methods  \cite{Beylkin91,Beylkin92,Beylkin93}, see Appendix~\ref{app1}.

\begin{algo}\label{algo6}
{ {\bf(off-line computation  of matrices  $\FF^{(\alpha)}$)} 
\begin{itemize} 
\item[i)] Compute  scaling function connection coefficients $$e^{\alpha}_{s_n}(\ell,\ell') \triangleq  {\color{black}(2\pi)^{-2\alpha}} \langle \varphi(2^{s_n}x-\ell), \left( \frac{-\partial^2}{\partial x^2}\right)^{\alpha}\varphi(2^{s_n}x-\ell' )\rangle_{\mathbb{R}}  $$ for any integers $\ell,\ell' $  by inversion of linear system \eqref{recursion} (see the explicit solution \eqref{eq:invSystem}),  where $\varphi$ denotes the scaling function associated to wavelet $\psi$.
\item[ii)] Construct the discrete $(2^{s_n})$-periodic function defined for any  integers $\ell,\ell'$ with $ 0 \le \ell,\ell' < 2^{s_n}$ by:
$$
 \begin{cases}
 e_{s_n}^{(\alpha)}(\ell,\ell') & \textrm{for} \hspace{0.3cm} 0\le |\ell' -\ell| < 2^{s_n-1}, \\
 e_{s_n}^{(\alpha)}(\ell,\ell'-2^{s_n}) & \textrm{for} \hspace{0.3cm} 2^{s_n-1} \le |\ell' -\ell| < 2^{s_n}.
\end{cases}
$$

\item[iii)] Approach $\FF^{(\alpha)}$ by performing the (discrete) FWT of   the latter two-dimensional  function multiplied by factor $2^{s_n}$ .\\
\end{itemize}
}
\end{algo}
Besides, note that for matrices $\FF^{(\alpha)}$ with  $\alpha \in \mathbb{N}$,  connection coefficients computed in step $i)$ of  algorithm \ref{algo6} are classically obtained by solving an eigenvalue problem, see \cite{Dahmen93}.
Using \eqref{eq:lowDimMatProdGrad}, we hereafter propose a fast  algorithm for the minimization problem  \eqref{objFunction6_}. \\

\begin{algo}\label{algo3}
{ {\bf(approach functional gradient w.r.t  $ d_n$)} 
\begin{itemize} 
\item[i)] Decompose $  (\bar y_1 -y_0 )\nabla    \bar y_1 $ by FWT using the dual divergence-free wavelet basis $\{\tilde \Psi_{{\boldsymbol \ell},{\boldsymbol s}};({{\boldsymbol \ell},{\boldsymbol s}})\in \Omega_n\}$  to obtain the  data-term gradient \eqref{gradJlDivFree},
\item[ii)]  Derive functional gradient  by adding the first terms of \eqref{eq:lowDimMatProdGrad} 
to  the data-term gradient.\\
\end{itemize}
}
\end{algo}
 The reconstruction algorithm is identical to algorithm \ref{algo4}. \\

\noindent
Algorithm \ref{algo3} can be substituted to  algorithm \ref{algo5} to approach the MAP estimate $ d^*_n$. In theory, it requires a 
theoretical complexity bounded by matrix multiplication, \ie $\mathcal{O}(n^3)$. In order to reduce this theoretical complexity, one may take advantage of the very sparse structure of matrices  $\FF^{(\alpha)}$, which could be investigated  as in \cite{Perrier99}. However, in practice FFT and FWT are the bottleneck of the computational cost. Algorithm \ref{algo3} requires only one  FWT  for each gradient step, in contrast to the numerous FFT and FWT involved in algorithm \ref{algo5}. All in all algorithm \ref{algo3} turns out to be  much faster  than  algorithm \ref{algo5}.\\

\section{Numerical evaluation}\label{sec:results}

In this section, we assess the performance of the  proposed estimation algorithms and compare them to recent state-of-the art methods. As explained below, the numerical evaluation relies on a synthetic benchmark of noisy image couples revealing the turbulence transport of a passive scalar. Turbulence is mimicked by synthesizing fBms for various Hurst exponents, which include  $H=\frac13$ and $H=1$ modeling respectively  3D  and 2D turbulence.    

\subsection{Divergence-free isotropic fBm fields generation}\label{sec:gen}

Divergence-free isotropic fBms were generated by a wavelet-based method relying on reconstruction formula \eqref{recUFourier}. More precisely, in agreement with the fBm model \eqref{priorMod},   the wavelets coefficients $\{\epsilon_ {{\boldsymbol \ell},{\boldsymbol s}}\}$ were sampled according to standard Gaussian white noise. The fBms realizations  were then synthesized  by application of algorithm~\ref{algo2}. 

\begin{figure}	[!ht]
\centerline{
\begin{tabular}{ccc}
\footnotesize{ $H=0.01$}&\includegraphics[width=0.32\textwidth]{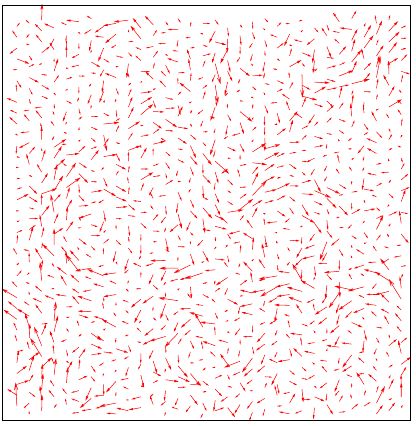}&\hspace{0.5cm}\includegraphics[width=0.32\textwidth]{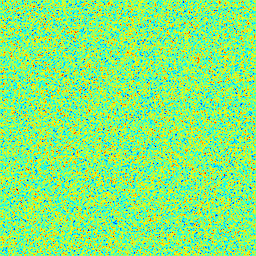}\\
\footnotesize{$H=\frac13$}&\includegraphics[width=0.32\textwidth]{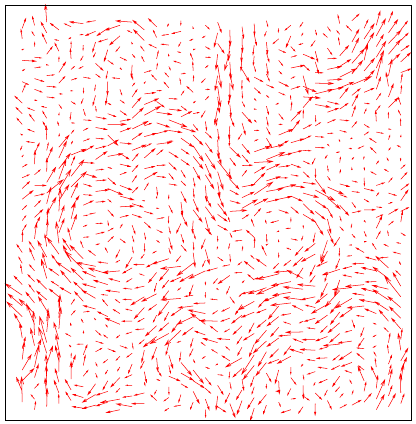}&\hspace{0.5cm}\includegraphics[width=0.32\textwidth]{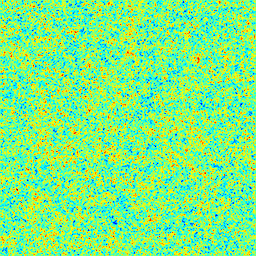}\\
\footnotesize{$H=\frac12$}&\includegraphics[width=0.32\textwidth]{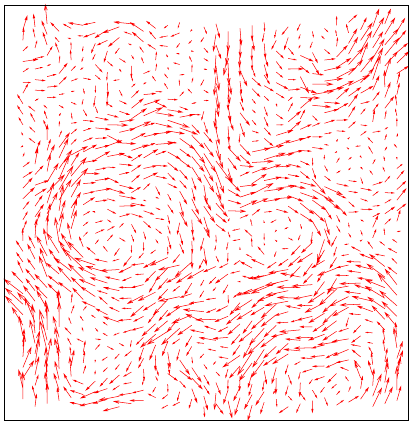}&\hspace{0.5cm}\includegraphics[width=0.32\textwidth]{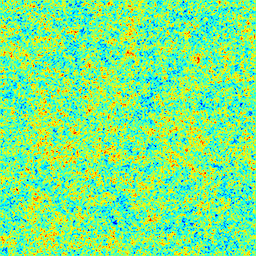}\\
\footnotesize{$H=\frac23$}&\includegraphics[width=0.32\textwidth]{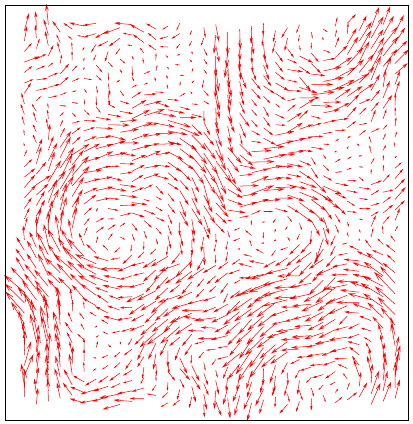}&\hspace{0.5cm}\includegraphics[width=0.32\textwidth]{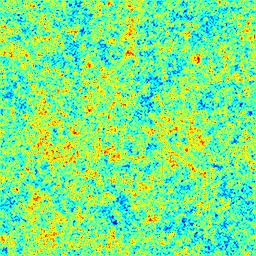}\\
\footnotesize{$H=1$}&\includegraphics[width=0.32\textwidth]{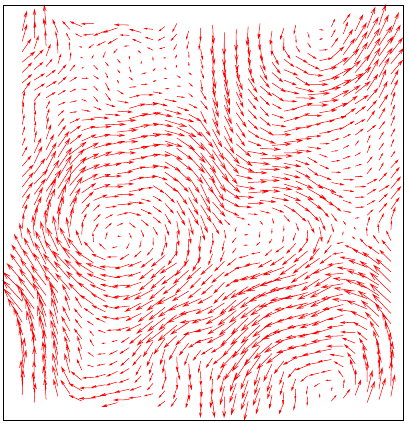}&\hspace{0.5cm}\includegraphics[width=0.32\textwidth]{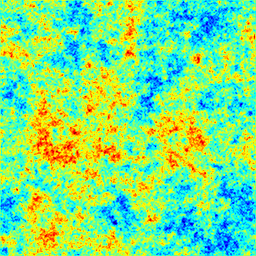}
\end{tabular}}
		\caption{{{\footnotesize  Synthesized fBms $\tvtr(\xx)=( u_1(\xx), u_2(\xx))^T$ for different values of $H$ (left). Associated vorticity maps, \ie  $\partial_x u_2(\xx)-\partial_y u_1(\xx)$  (right) \label{fig:uvGroundTruth}\label{fig:vortGroundTruth}}}. }\vspace{-0.cm}
\end{figure}

 In order  to form an evaluation benchmark for the regularization model, a set of 5 fBm realizations  were synthesized (from an identical white noise realization) with a resolution  $2^{8} \times 2^{8}$ on the domain $[0,1]^2$. The wavelet generator was constructed from periodized Coiflets with 10 vanishing moments, so-called Coiflets-10~\cite{mallat2008wavelet}. Note that these wavelets are 3 times differentiable so that our regularity assumptions are satisfied for any $H \le 1$.   The realizations are  associated to  Hurst exponents  $H_1=0.01$, $H_2 =\frac{1}{3}$, $H_3 =\frac{1}{2}$, $H_4 =\frac{2}{3}$ and $H_5=1$. 
Let us remark that  the cases $H=\frac13$ and $H=1$ are consistent with isotropic turbulence models introduced by Kolmogorov for 3D flows \cite{Kolmogorov41} (resp. by Kraichnan for 2D flows  \cite{Kraichnan67}), while the case $H=\frac12$ constitutes  a standard Brownian motion.  
Parameter $\sigma$  is chosen so that for any $\xx$ belonging to the pixel grid, the displacement $|\mathbf{u}(\xx)|$ is not greater that  $10 \times 2^{-8}$, \ie 10 pixels.
Figure~\ref{fig:uvGroundTruth}  displays  vector fields $\tvtr(\xx)= (u_1(\xx), u_2(\xx))^t$ and vorticity maps $\partial_x u_2(\xx)-\partial_y u_1(\xx)$ associated to the 5 fBms. 

Let the radial power spectrum of ${\tvtr}$ be defined by: 
\begin{align}\label{eq:isoSpec}
E(\kappa)\triangleq \int_{\mathcal{S}_\kappa} E_1(\xx)+E_2(\xx) d\xx,
\end{align}
 where ${\mathcal{S}_\kappa}$ is the circle of radius $\kappa$.
It is easy to check that  according to \eqref{powerspectrum}, this function is $E(\kappa)
=c\, \kappa^{-2H-1}$, where $c>0$. Therefore, the  spectra of our 5 different fBms  decay exponentially with power  respectively equal  to $-1.02$, $-\frac{5}{3}$, $-2$, $-\frac{7}{3}$ and $-3$. 

\subsection{Synthetic image couple generation}

\begin{figure}	[!ht]
\begin{tabular}{cccc}
\footnotesize{$H=\frac13$}&\includegraphics[width=0.32\textwidth]{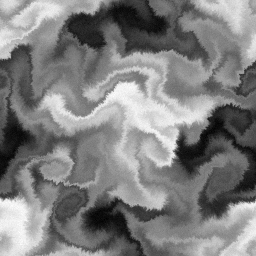}&\includegraphics[width=0.32\textwidth]{UVtZ-1_66667}&\includegraphics[width=0.32\textwidth]{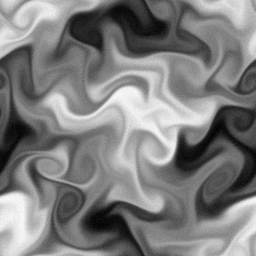}\\
\footnotesize{$H=1$}&\includegraphics[width=0.32\textwidth]{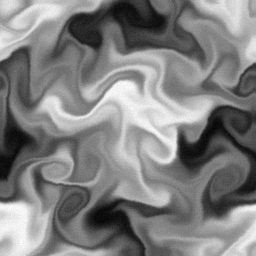}&\includegraphics[width=0.32\textwidth]{UVtZ-3}&\includegraphics[width=0.32\textwidth]{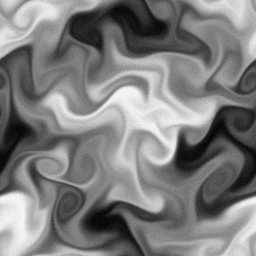}
\end{tabular}
		\caption{ {\footnotesize  Initial image $y_0$ (left), deformation field $\tvt$ (middle) and final image  $y_1$ (right) for $H=\frac13$ and $H=1$. Images $y_0$ and $y_1$ have been  corrupted by a white noise. \label{fig:Im} }}\vspace{-0.cm}
\end{figure}

 To simulate the couple $(y_0,y_1)$ in the data-term \eqref{objFunction}, we start by  a fixed image $y_1$ and derive $y_0$ from the relation $y_0(\xx) =y_1(\xx+ \tvtr(\xx))$ derived in  section \ref{sec:likeli}.  Since $\xx+\tvt(\xx)$ does not necessarily lie on the pixel grid,   we used cubic B-splines for interpolation of $y_1$.  
 To simulate realistic measurement conditions, the so-generated  images were then corrupted by i.i.d. Gaussian noise yielding a peak signal to noise ratio (PSNR) on $y_0$ (resp. $y_1$) of 33.2 dB (resp. 33.5 dB).  The resulting  image pairs are displayed in figure~\ref{fig:Im} for $H=\frac13$ and $H=1$.

\subsection{Optic-flow evaluation procedure}\label{sec:refmethods}

 The  divergence-free fBm fields were estimated according to a MAP criterion, solving  minimization problems \eqref{objFunction6} or \eqref{objFunction6_}.  The proposed approaches are compared to two other standard regularizers, which all require the choice of a basis to decompose $\tvt$.   In order to make relevant comparisons,  we chose the divergence-free wavelet basis for all these alternatives.  The wavelet generator was constructed from divergence-free biorthogonal Coiflets-10  with periodic boundary conditions.  
Moreover, the optimization procedure used for all regularizers was the same and relied on an identical  data-term.

   The five different estimation methods  used for evaluation  are listed and detailed hereafter. They are denoted A, B, C, D and E. Methods A and B are state-of-the-art algorithms while methods C, D and E implement the fBm prior.
 \begin{itemize}
\item[A -] \textit{Penalization of $L^2$ norm of velocity components gradients}. The most common  approach in optic-flow estimation, as first proposed in \cite{Horn81}, is to penalize the $L^2$ norm of  the velocity gradients. We used the wavelet-based implementation proposed in \cite{KadriHarouna12}. 
\item[B -] \textit{Penalization of $L^2$ norm of vorticity gradient}. In fluid motion estimation a popular approach  is to penalize the $L^2$ norm of the vorticity gradient  \cite{Corpetti02,Suter94,Yuan07}.    Here again, we used the wavelet-based implementation proposed in \cite{KadriHarouna12}.
\item[C -]  \textit{fBm regularization  in a fractional Laplacian wavelet basis}. This corresponds to solve  \eqref{objFunction6} using algorithms \ref{algo1} and \ref{algo2}. 
\item[D -]  \textit{fBm regularization   in a divergence-free wavelet basis}. This corresponds to  solve  \eqref{objFunction6_} computed without approximation using algorithms \ref{algo5} and \ref{algo4}.
\item[E -]  \textit{Approached fBm regularization  in a divergence-free wavelet basis}. This corresponds to solve \eqref{objFunction6_} where the regularization term is approached  using algorithms \ref{algo4} and \ref{algo3}.

\end{itemize}
Each of the regularizer in methods A, B, C, D and E were optimally tuned, that is to say  regularization coefficient  were chosen (using a brute-force approach) in order to the RMSE detailed hereafter.  
 Note that an implicit regularization by  polynomial approximation has also been tested. It  is a well-known approach in computer vision \cite{Becker11,Derian12wavelet,Lucas81,Wu00}. The performances were clearly below the previous approaches, so we do not display the results in this paper.

 Let $\mathcal{S}$ denote the set of pixel sites. The two following  criteria were used to evaluate the  accuracy of estimated fields denoted by ${\tvtr}^* $: the Root Mean Squared end-point Error (RMSE) in pixel
\begin{eqnarray*}
&&\text{RMSE}= \left( \frac{1}{n^2} \sum_{\xx \in \mathcal{S}} \big|\tvtr^*(\xx) - \tvtr_{\rm }(\xx)  \big|^2 \right)^{\frac{1}{2}} ,
\end{eqnarray*}
and the Mean Barron Angular Error (MBAE) in degrees
\begin{eqnarray*}
&&\text{MBAE} = \frac{1}{n^2} \sum_{\xx  \in \mathcal{S}}  \text{arcos}\left( \frac{\tvtr^*(\xx) \cdot \tvtr_{\rm }(\xx)}{|\tvtr{\rm }(\xx)|^2}\right), 
\end{eqnarray*}
where  $\tvtr$ represents the synthesized fBm. Moreover, we introduce a criterion to evaluate the accuracy of  reconstruction of the  power-law decay of the radial power spectrum. More precisely, in logarithmic coordinates  the power-law  \eqref{eq:isoSpec} writes as an affine function of $\log(\kappa)$ of the form $ -(2H+1)\log(\kappa)+\gamma$, depending on two parameters, namely the Hurst exponent $H$ and the intercept  $\gamma$, that can be explicitly related to $\sigma$ in \eqref{powerspectrum}.   Performing a  linear regression (by the ordinary least squares method) on the estimated spectrum in logarithmic coordinates, we obtain an estimation of the affine function parameters  denoted by $ H^*$ and $ \gamma^*$. The quality criterion is then chosen to be the $L^1$ distance between the estimated and true affine functions within the interval $[\log(\kappa_{min}),\log(\kappa_{max})]$ , called  the   Spectrum  Absolute Error (SAE):

\begin{eqnarray*}
&&\text{SAE} =\int_{\log(\kappa_{min})}^{\log(\kappa_{max})} \!|2x(H- H^*) + \gamma^*-\gamma |dx, 
\end{eqnarray*}

In order to evaluate the power-law reconstruction at small scales, we chose $\kappa_{min}=10$ and $\kappa_{max}=n$.

Finally, we performed an additional visual comparison of the accuracy of restituted vorticity maps.

\subsection{Results}

\begin{figure}	[!ht]
\hspace{-1cm}\begin{tabular}{c|c}
\footnotesize{$H=\frac13$}&\footnotesize{$H=1$}\\
\includegraphics[width=0.65\textwidth]{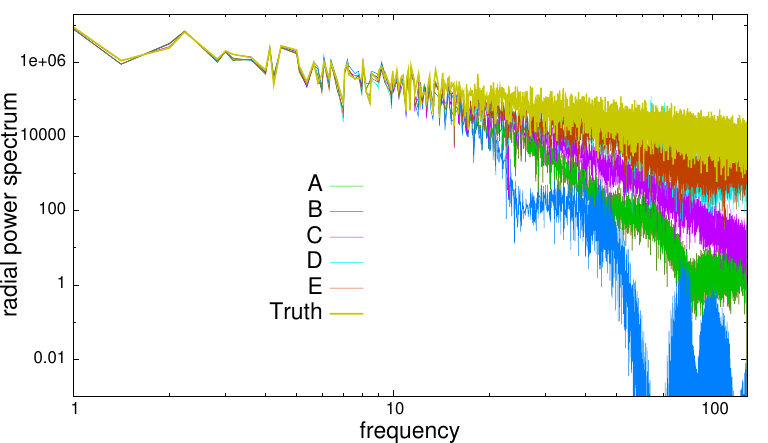}&\includegraphics[width=0.65\textwidth]{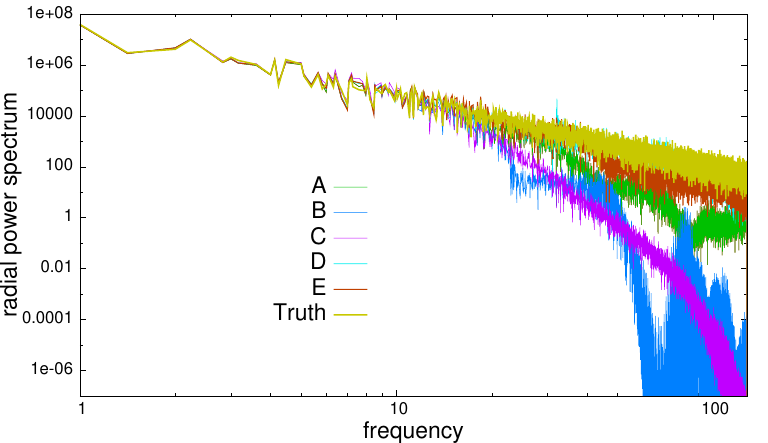}\\
\includegraphics[width=0.65\textwidth]{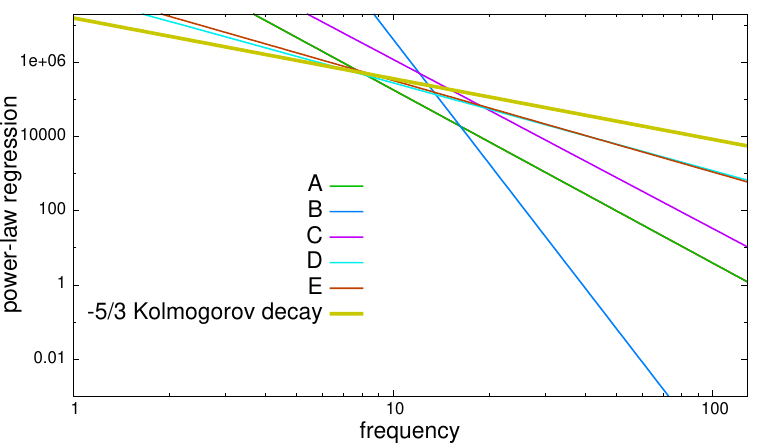}&\includegraphics[width=0.65\textwidth]{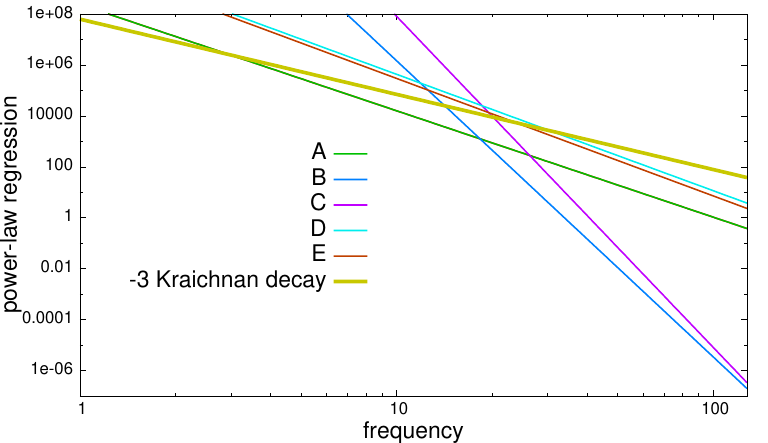}\\
\vspace{-0.25cm}
\end{tabular}
		\caption{{\footnotesize Radial power spectrum estimates in logarithmic coordinates (above) and ordinary least squares  fitting (below) for methods A to E (see section \ref{sec:refmethods} for a description) for $H=\frac13$ (left) and $H=1$ (right), compared to ground truth and the theoretical decay.    Regularization coefficients where  chosen to minimize the RMSE.  \label{fig:SpectrumEstim} }}\vspace{-0.cm}
\end{figure}

\begin{figure}	[!ht]
\hspace{1cm}\begin{tabular}{ccc}
\multicolumn{3}{c}{{\footnotesize $H=\frac13$}}\\
\hline \\
\includegraphics[width=0.32\textwidth]{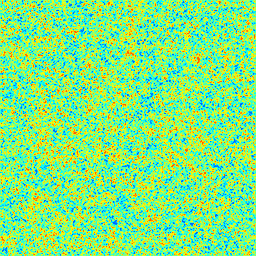}&\includegraphics[width=0.32\textwidth]{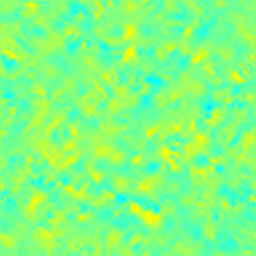}&\includegraphics[width=0.32\textwidth]{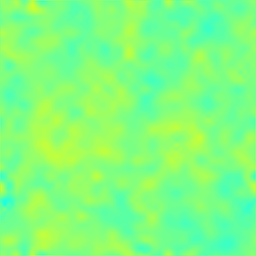}\vspace{-0.15cm}\\
\footnotesize{Truth}&\footnotesize{A}&\footnotesize{B}\vspace{0.1cm}\\
\includegraphics[width=0.32\textwidth]{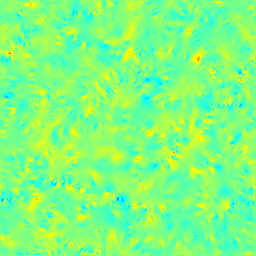}&\includegraphics[width=0.32\textwidth]{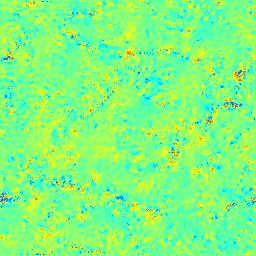}&\includegraphics[width=0.32\textwidth]{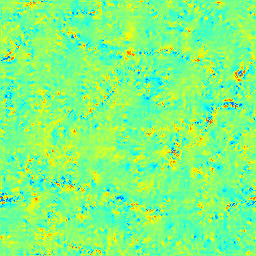}\vspace{-0.15cm}\\
\footnotesize{C}&\footnotesize{D}&\footnotesize{E}\vspace{0.1cm}\\
\multicolumn{3}{c}{{\footnotesize $H=1$}}\\
\hline \\
\includegraphics[width=0.32\textwidth]{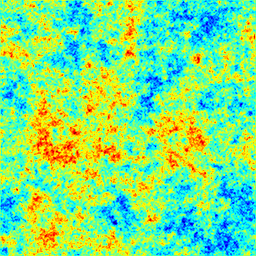}&\includegraphics[width=0.32\textwidth]{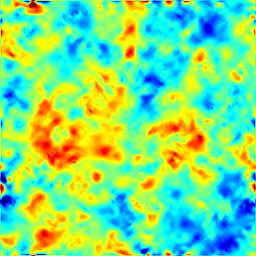}&\includegraphics[width=0.32\textwidth]{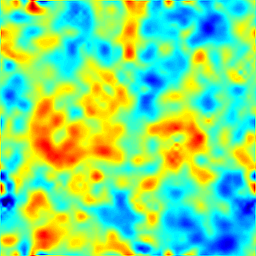}\vspace{-0.15cm}\\
\footnotesize{Truth}&\footnotesize{A}&\footnotesize{B}\vspace{0.1cm}\\
\includegraphics[width=0.32\textwidth]{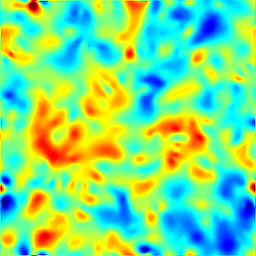}&\includegraphics[width=0.32\textwidth]{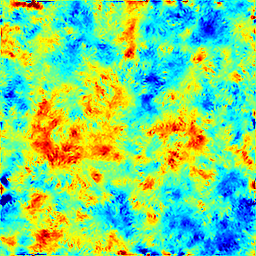}&\includegraphics[width=0.32\textwidth]{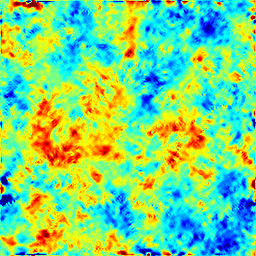}\vspace{-0.15cm}\\
\footnotesize{C}&\footnotesize{D}&\footnotesize{E}\\
\vspace{-0.32cm}
\end{tabular}
		\caption{{\footnotesize  Estimated vorticity maps for methods A to E (see section \ref{sec:refmethods} for a description) and for $H=\frac13$ (above) and $H=1$ (below). Regularization coefficients were  chosen to minimize the RMSE  \label{fig:vortEstimh}}}\vspace{-0.cm}
\end{figure}

In table of figure~\ref{fig:performance}, the performance of the proposed methods (C, D and E) can be compared in terms of RMSE, MBAE and SAE to state-of-the-art approaches (A and B). Let us comment  these results. 
Methods C, D and E yield the best results with respect to each criterion. Method C, \ie the method based on fractional Laplacian wavelets, provides the lowest RMSE and MBAE.  An average RMSE gain of 19\%  with respect to the best state-of-the-art method is observed, with a peak at 26\% for $H=\frac12$. However, considering the 3 criteria jointly, methods D and E, \ie exact and approached method based on divergence-free wavelets, provide the best compromise. In particular, according to SAE it can be noticed that,  unlike method C,  methods D and E achieve to accurately reconstruct the  power-law decay of the fBm spectrum. This is illustrated in figure \ref{fig:SpectrumEstim}. Moreover, the approximation used to derive method E seems to be accurate since performance of E are very close to those of  D.  In figure \ref{fig:vortEstimh}, one can visualize estimated vorticity maps with the different methods for $H=\frac13$ and $H=1$, \ie fBms modeling respectively  3D or 2D turbulence.  This figure clearly shows the superiority of methods D and E in reconstructing the fractal structure of the vorticity fields. 

\begin{figure}	[!ht]
\hspace{-1cm}\begin{tabular}{c|c}
\footnotesize{$H=\frac13$}&\footnotesize{$H=1$}\\
\includegraphics[width=0.65\textwidth]{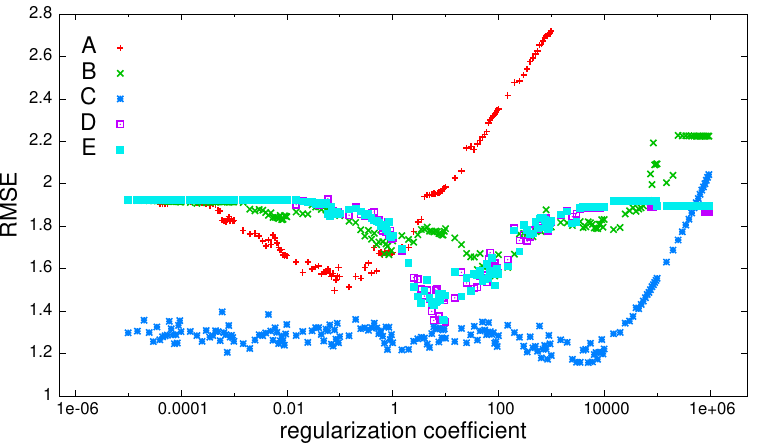}&\includegraphics[width=0.65\textwidth]{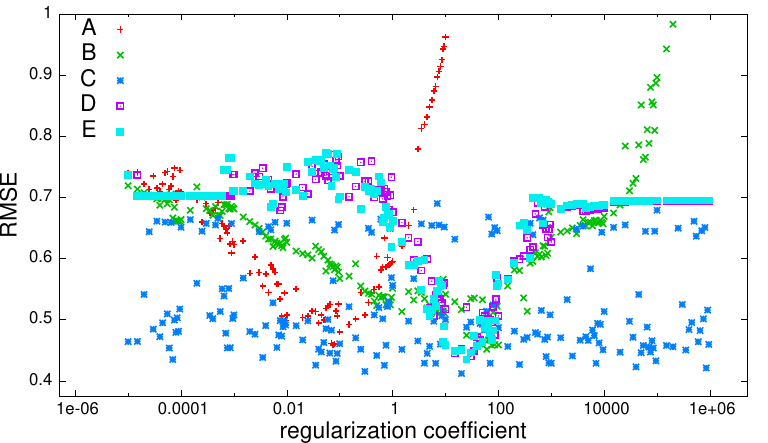}\\
\includegraphics[width=0.65\textwidth]{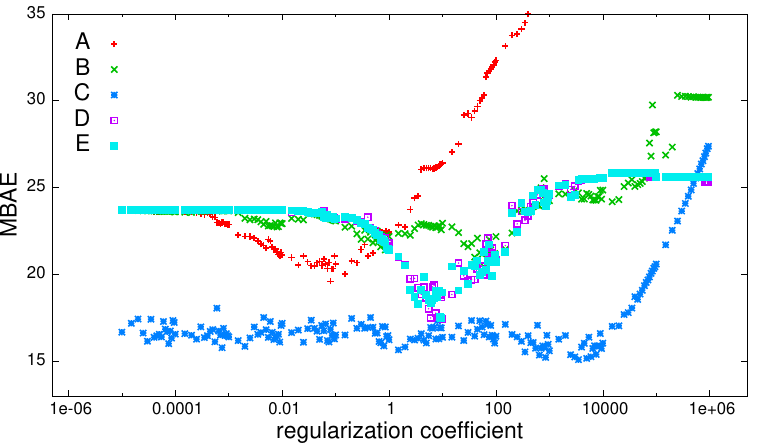}&\includegraphics[width=0.65\textwidth]{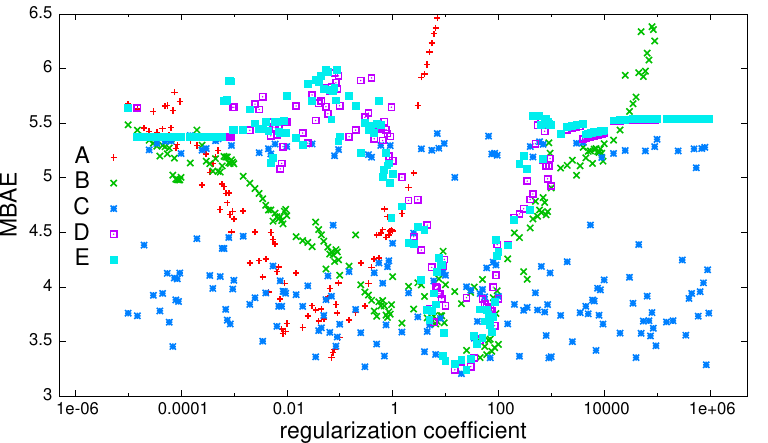}\\
\includegraphics[width=0.65\textwidth]{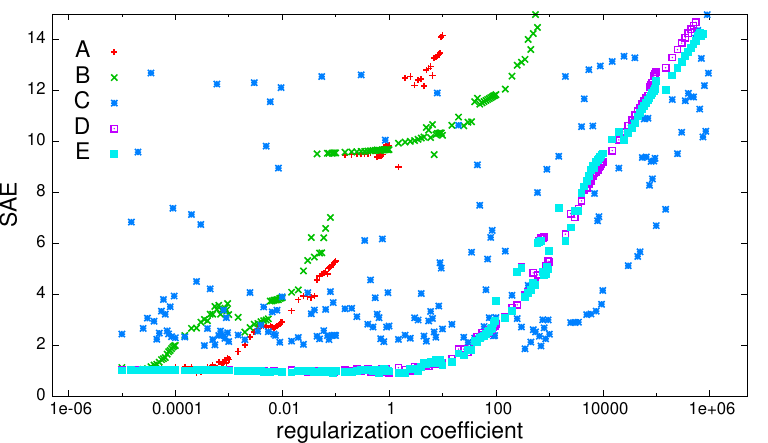}&\includegraphics[width=0.65\textwidth]{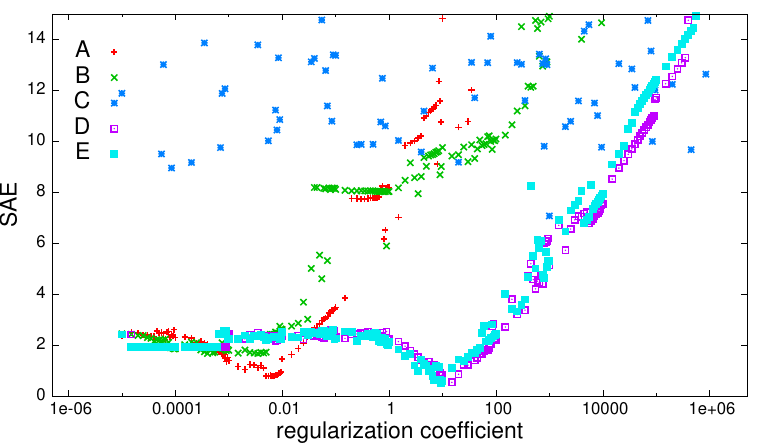}\\
\vspace{-0.25cm}
\end{tabular}
		\caption{{\footnotesize Influence of regularization coefficient on the accuracy of the estimate in terms of RMSE (above), MBAE (middle) and SAE (below) 
		for methods A to E (see section \ref{sec:refmethods} for a description) and for $H=\frac13$ (left) and $H=1$ (right). \label{fig:errorFctAlpha} }}\vspace{-0.cm}
\end{figure}

\begin{figure}	[!ht]
\hspace{1cm}\begin{tabular}{ccc}
\multicolumn{3}{c}{{\footnotesize $H=\frac13$}}\\
\hline \\
\includegraphics[width=0.32\textwidth]{vortfBmZ-1_66667__.png}&\includegraphics[width=0.32\textwidth]{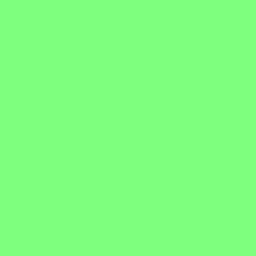}&\includegraphics[width=0.32\textwidth]{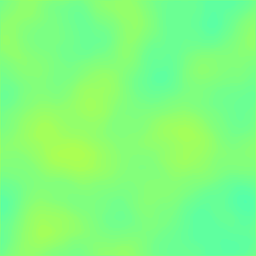}\vspace{-0.15cm}\\
\footnotesize{Truth}&\footnotesize{A}&\footnotesize{B}\\
\includegraphics[width=0.32\textwidth]{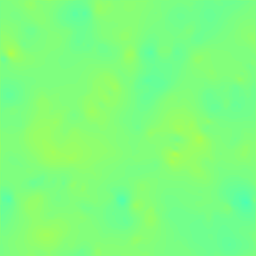}&\includegraphics[width=0.32\textwidth]{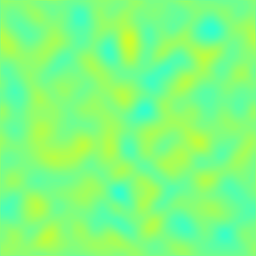}&\includegraphics[width=0.32\textwidth]{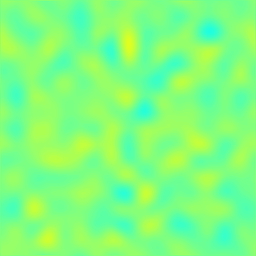}\vspace{-0.15cm}\\
\footnotesize{C}&\footnotesize{D}&\footnotesize{E}\\
\multicolumn{3}{c}{{\footnotesize $H=1$}}\\
\hline \\
\includegraphics[width=0.32\textwidth]{vortfBmZ-3_.png}&\includegraphics[width=0.32\textwidth]{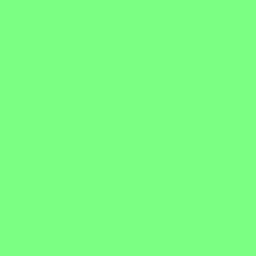}&\includegraphics[width=0.32\textwidth]{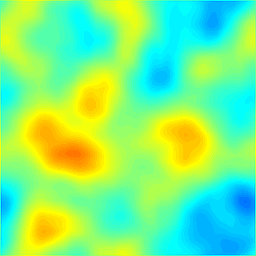}\vspace{-0.15cm}\\
\footnotesize{Truth}&\footnotesize{A}&\footnotesize{B}\\
\includegraphics[width=0.32\textwidth]{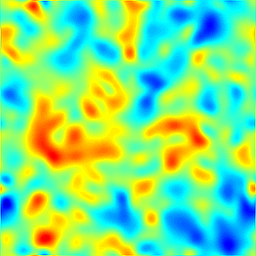}&\includegraphics[width=0.32\textwidth]{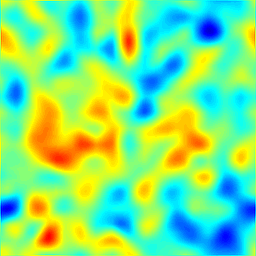}&\includegraphics[width=0.32\textwidth]{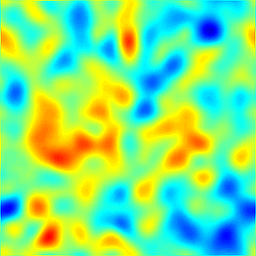}\vspace{-0.15cm}\\
\footnotesize{C}&\footnotesize{D}&\footnotesize{E}\\
\vspace{-0.25cm}
\end{tabular}
		\caption{{\footnotesize  Estimated vorticity maps for methods A to E  (see section \ref{sec:refmethods} for a description) for  $H=\frac13$ and $H=1$  in the case of a  very large regularization coefficient  $(\sim1e6)$ \label{fig:vortEstimRegLarge}}}\vspace{-0.cm}
\end{figure}

Plots of figure \ref{fig:errorFctAlpha} show the influence of the regularization parameter in terms of RMSE, MBAE and SAE for $H=\frac13$ and $H=1$.   Clear minima of RMSE and MBAE are visible for methods A, B, D and E. On the other hand, method C, \ie  fractional Laplacian wavelet basis, seems to be `unstable' in the sense that it yields inhomogeneous performances for small variations of regularization parameter. The saturation of the RMSE and the MBAE for large values of the regularization parameter shows that, on the contrary to state-of-the-art methods, sensitivity of method D and E to the latter parameter is weak, in the sense that it yields reasonable estimation error even for regularization parameter far from an optimal value. This error saturation effect is illustrated in  figure \ref{fig:vortEstimRegLarge}. It displays  vorticity maps produced by the different methods for a very large regularization coefficient. 

\section{Conclusion}
This work addresses  the inverse-problem of estimating a hidden turbulent motion field from the observation of a pair of images.
We adopt a Bayesian framework where we propose a family of divergence-free, isotropic, self-similar priors for this hidden field. Self-similarity and divergence-free are well known features of incompressible turbulence in statistical fluid mechanics.  
Our priors are bivariate fractional Brownian fields, resulting  from the extra  assumptions that the hidden field is  Gaussian and has  stationary increments.
The main purpose of this article is the design of effective and efficient algorithms to achieve  MAP estimation, by  expanding these specific priors into well-chosen bases. 
From a spectral integral representation proved in Proposition~\ref{prop:spectral},   we represent divergence-free fBms in two specific wavelet bases.   The first option (Proposition~\ref{corrol01}) is a fractional Laplacian wavelet basis which plays the role of a whitening filter  in the sense that the wavelet coefficients are uncorrelated.   The second alternative  is to use a divergence-free wavelet basis, which is well-suited to our case. The latter wavelets simplify the decomposition, since  they neither involve fractional  operators nor Leray projector on the divergence-free functional space. However, the wavelets coefficients are then correlated. We provide a closed-form expression for the induced correlation structure (Proposition~\ref{prop:covDls}), which  is necessary to implement this second approach in practice. For these two approaches, the algorithms to reach the MAP involve gradient based LBFGS optimization procedures and rely on fast transforms (FFT or/and FWT).
Moreover we propose an approximation of the  correlation structure of the coefficients in the divergence-free wavelets expansion. It is  based on off-line computation of fractional Laplacian wavelet connection coefficients. This approximation leads to the fastest algorithm without loss of accuracy.
According to an intensive  numerical evaluation carried out in section \ref{sec:results}, all proposed algorithms clearly  outperform the state-of-the-art methods. 
Finally, in the light of our experiments, the divergence-free wavelet expansion seems  to be the most appropriated representation to solve our MAP inverse-problem.

An obvious and important perspective is the assessment of the develops algorithms in the context of  real turbulence. 
To simplify the exposition, this work essentially focuses on the bi-variate case, which is of interest in particular geophysical contexts. However, there may be some limitations in studying  three-dimensional turbulence from bi-dimensional slices or projections of the flow \cite{Frisch95}. Acquisition of three-dimensional data is not an easy task. In fact  volume data is generally reconstructed from bi-dimensional information and this inverse problem still represents an active domain of research.
Nevertheless,  extension of our algorithms to the three-dimensional case is straightforward since no theoretical or technical issue constitute a block. 

%
\section*{Acknowledgements}

The authors wish to acknowledge Pierre D\'erian for fruitful discussions on wavelets and their implementation. They are also sincerely grateful to anonymous referees for their numerous insightful comments and suggestions which considerably helped them in improving the first version of the paper.\\

\appendix
\section{Proofs}

\subsection{Proof of  Proposition~\ref{prop:spectral}}\label{proofofprop:spectral}

Let us recall (see e.g. \cite{Yaglom:87}) that given a standard Gaussian spectral measure $\tilde W_1$, the integral $\int f(\kk) \tilde W_1(d\kk)$ is well-defined whenever  $f\in L^2(\mathbb R^2)$, has zero expectation and for  $f, g$ in $L^2(\mathbb R^2)$: 
\begin{equation}\label{corint} \mathbb{E}\left(\int f(\kk) \tilde W_1(d\kk)\overline{\int g(\kk) \tilde W_1(d\kk)}\right)=\int f(\kk) \overline{g(\kk)} d\kk.\end{equation}

In \eqref{spectral}, the matrix ${\bf P}(\kk)\triangleq \left[ \mathbf{I} - \frac{\kk\kk^T}{ \| {\kk }\|^2}\right]$ corresponds to the Leray projection matrix in the Fourier domain. It is easily verified that all entries of ${\bf P}$ are in $[0,1]$. For this reason the integral \eqref{spectral} is well defined, since for all $\xx\in\mathbb{R}^2$ and all $H\in(0,1)$,  the function $\kk\mapsto (e^{i{\bf k} \cdot \xx} -1)\|\kk\|^{-H-1}$ belongs to $L^2(\mathbb{R}^2)$.

Let us show that the structure function of $\tvt$ is given by \eqref{eq:crosscorrel}. 
For $j=1,2$, denote ${\bf e}_j$  the bivariate vector whose $j$-th  component is  equal to one while the other component is zero.
For all $i,j=1,2$, from \eqref{spectral} and \eqref{corint}, since ${\bf P}^2={\bf P}$, we get
 \begin{align*} 
& \mathbb{E}[(\tvt_i(\xx_2)-\tvt_i(\xx_1))\overline{(\tvt_j(\xx_4)-\tvt_j(\xx_3))}] \\
 &= \frac{\sigma^2}{(2\pi)^2} \int_{\mathbb{R}^2}\ \| \kk \|^{-2H-2} {\bf e}_i^T {\bf P}(\kk) {\bf e}_j \left(e^{i\kk \cdot \xx_2}-e^{i\kk \cdot \xx_1}\right)\left(e^{- i\kk \cdot \xx_4}- e^{-i\kk \cdot \xx_3}\right)d\kk \\
 & = \frac{\sigma^2}{(2\pi)^2} \int_{\mathbb{R}^2}\ \| \kk \|^{-2H-2} {\bf e}_i^T {\bf P}(\kk) {\bf e}_j \left(e^{i\kk \cdot (\xx_2-\xx_4)}-e^{i\kk \cdot (\xx_2-\xx_3)}-e^{i\kk \cdot (\xx_1-\xx_4)}+e^{i\kk \cdot (\xx_1-\xx_3)}\right)d\kk.
  \end{align*}

We use Lemma 2.2 in \cite{Tafti10} to get the following Fourier transform: for any $\xx\in\mathbb R^2$
$$\frac{1}{(2\pi)^2}\int_{\mathbb{R}^2}  \| \kk \|^{-2H-2} {\bf P} (\kk) e^{i\kk \cdot \xx} d\kk =c_H\|\xx\|^{2H} \left(2H \frac{\xx\xx^T}{\|\xx\|^2} - (2H+1)\mathbf{I} \right).$$
where $c_H= {\Gamma(1-H)}/(\pi 2^{2H+2} \Gamma(H+1) H(2H+2))$. The structure function  \eqref{eq:crosscorrel} is then  deduced.

Finally  \eqref{eq:crosscorrel} coincides with the structure function obtained in \cite{Tafti10} Section 4.5 when $\tvt$ is defined by \eqref{sol_poisson}. As this structure function characterizes the law of the Gaussian vector field $\tvt$, this shows that the two vector fields defined by \eqref{spectral} and   \eqref{sol_poisson}  share the same distribution.

  \begin{flushright}{$\square$}\end{flushright}

\subsection{Proof of  Proposition~\ref{corrol01}}\label{proofofcorrol01}

Let us denote by $\Phi_0$ the indicator function  $\mathbb{I}_{[0,1]^2}(\xx)$, so that according to the construction explained in Section \ref{sec:wave}, the wavelets $\Phi_{{\boldsymbol \ell},{\boldsymbol s}}$, for $({\boldsymbol \ell},{\boldsymbol s})\in\Omega\cup 0$, form an orthonormal basis of  $  L^2([0,1]^2)$.
For any function $\wwr \in L^2([0,1]^2)$, we have  $\wwr (\xx)= \sum_{({\boldsymbol \ell},{\boldsymbol s})\in\Omega\cup 0}\langle \wwr,  {\Phi}_{{\boldsymbol \ell},{\boldsymbol s}} \rangle \Phi_{{\boldsymbol \ell},{\boldsymbol s}} (\xx)$ in $L^2([0,1]^2)$, where $\langle .,.\rangle$ denotes the scalar product in $L^2([0,1]^2)$. \textcolor{black}{The same relation holds in  $L^2(\mathbb R^2)$ if each function is extended outside $[0,1]^2$ by zero, and we denote these extensions $\ww^0$ and $ {\Phi}^0_{{\boldsymbol \ell},{\boldsymbol s}}$ respectively. } Hence, by the Plancherel's theorem, we deduce that   $\mathcal F \ww^0({\bf k}) = \sum_{\ell, {\boldsymbol s}} \langle \mathcal F\ww^0, \mathcal F{\Phi}^0_{{\boldsymbol \ell},{\boldsymbol s}} \rangle \mathcal F { \Phi}^0_{{\boldsymbol \ell},{\boldsymbol s}} ({\bf k})$ in $L^2(\mathbb R^2)$, where now $\langle .,.\rangle$ denotes the scalar product in $L^2(\mathbb R^2)$.
Therefore, for $j =1,2$,
\begin{align}\label{eq:decDW}
\int_{\mathbb{R}^2}  \mathcal F\ww^0({\bf k}) \tilde W_j (d {\bf k})&
= \sum_{\ell, {\boldsymbol s}}  \langle \mathcal F\ww^0, \mathcal F{\Phi}^0_{{\boldsymbol \ell},{\boldsymbol s}} \rangle \eta^j_ {{\boldsymbol \ell},{\boldsymbol s}}
\end{align} 
with
\begin{align}\label{WienerIntPhi}
\eta^j_ {{\boldsymbol \ell},{\boldsymbol s}} =\int_{\mathbb{R}^2} \mathcal F{ \Phi}^0_{{\boldsymbol \ell},{\boldsymbol s}} ({\bf k}) \tilde W_j (d {\bf k}).
\end{align}

From  \eqref{corint} and the Plancherel's theorem, since the  wavelets are orthogonal and normalized in $L^2$, we note that $\eta^j_ {{\boldsymbol \ell},{\boldsymbol s}}$ are i.i.d standard Gaussian random variables. 

\textcolor{black} {Now recall from \eqref{spectral} that 
$\tvt (\xx) = (\tvt_1(\xx),\tvt_2(\xx) )^T$ where for $i=1,2$, $\tvt_i(\xx)=\int \mathcal F \ww^0_{i1}(\kk)\tilde W_1(d\kk)+ \int \mathcal F \ww^0_{i2}(\kk)\tilde W_2(d\kk)$ with 
\begin{equation}\label{refere1}
\mathcal F \ww^0_{ij}(\kk)=\frac{\sigma}{2\pi}\, (e^{i{\bf k} \cdot \xx} -1)  \|\kk\|^{-H-1} (\delta_{ij} -  {k}_i{ k}_j/ \| {\bf k \| ^2 }).\end{equation}
Applying \eqref{eq:decDW} to $\wwr=\ww_{i1},\ww_{i2},$ we obtain for $i=1,2$
\begin{align*}\tvt_i(\xx) & = \sum_{({\boldsymbol \ell},{\boldsymbol s})\in\Omega\cup 0,\, j\in \{1,2\}} \eta^j_ {{\boldsymbol \ell},{\boldsymbol s}}\langle  \mathcal F \ww^0_{ij}, \mathcal F{\Phi}^0_{{\boldsymbol \ell},{\boldsymbol s}}\rangle\\
& =\sum_{({\boldsymbol \ell},{\boldsymbol s})\in\Omega\cup 0,\, j\in \{1,2\}} \eta^j_ {{\boldsymbol \ell},{\boldsymbol s}} \int_{\mathbb{R}^2}  \mathcal F\ww^0_{ij}(\kk) \mathcal F{\Phi}^0_{{\boldsymbol \ell},{\boldsymbol s}}(\kk)d\kk,
\end{align*} 
and from \eqref{refere1} we deduce  the representation} 
\begin{align}\label{WienerInt3bis}
\tvt(\xx)&=\frac{\sigma}{2\pi}\sum_{({\boldsymbol \ell},{\boldsymbol s})\in\Omega\cup 0} \int_{\mathbb{R}^2} (e^{i{\bf k} \cdot \xx} -1)   \left[ \mathbf{I} - \frac{\kk\kk^T}{ \| {\kk }\|^2}\right] \boldsymbol\eta_ {{\boldsymbol \ell},{\boldsymbol s}} \|\kk \|^{-H-1}  \mathcal F{\Phi}^0_{{\boldsymbol \ell},{\boldsymbol s}}({\bf k})d {\bf k},
 \end{align} 
where $\boldsymbol\eta_{{\boldsymbol \ell},{\boldsymbol s}}\triangleq(\eta^1_{{\boldsymbol \ell},{\boldsymbol s}},\eta^2_{{\boldsymbol \ell},{\boldsymbol s}})^T$. 

Since the mother wavelet $\psi$ has $M$ vanishing moments, for any $({\boldsymbol \ell},{\boldsymbol s})\in\Omega$, $\Phi_{{\boldsymbol \ell},{\boldsymbol s}}$ has $M$ vanishing moments along at least one direction (say $\xx_1$). As a consequence, there exists a bounded function ${\theta}_{{\boldsymbol \ell},{\boldsymbol s}}$ such that $\mathcal F{\Phi}^0_{{\boldsymbol \ell},{\boldsymbol s}}(\kk)=(-ik_1)^M{\theta}_{{\boldsymbol \ell},{\boldsymbol s}}(\kk)$  (see \cite{mallat2008wavelet}). So  $|\|\kk \|^{-H-1}  \mathcal F{\Phi}^0_{{\boldsymbol \ell},{\boldsymbol s}}({\bf k})|=|\|\kk \|^{-H-1}(-ik_1)^M {\theta}_{{\boldsymbol \ell},{\boldsymbol s}}(\kk)|\leq c\|\kk \|^{M-H-1}$, where $c$ is some positive constant. Since $M>H$, the latter bound shows that $\kk\mapsto\|\kk \|^{-H-1}  \mathcal F{\Phi}^0_{{\boldsymbol \ell},{\boldsymbol s}}({\bf k})$ is square-integrable on any compact. Moreover it is square-integrable at infinity as $\kk\mapsto \mathcal F{\Phi}^0_{{\boldsymbol \ell},{\boldsymbol s}}({\bf k})$ is, while $\kk\mapsto\|\kk \|^{-H-1}$ asymptotically vanishes. Hence, for any $({\boldsymbol \ell},{\boldsymbol s})\in\Omega$, $\kk\mapsto\|\kk \|^{-H-1}  \mathcal F{\Phi}^0_{{\boldsymbol \ell},{\boldsymbol s}}({\bf k})\in L^2(\mathbb{R}^2)$ and the integral in \eqref{WienerInt3bis} can be split, leading to the representation in  $(L^2([0,1]^2))^2$,
\begin{align}\label{wave2}
\tvt(\xx)=2\pi\sigma \sum_{({\boldsymbol \ell},{\boldsymbol s})\in \Omega} \, & \mathcal{P} \left[ \boldsymbol\eta_{{\boldsymbol \ell},{\boldsymbol s}} {\Phi}^{(-H-1)}_{{\boldsymbol \ell},{\boldsymbol s}}\right]({\xx})  - \mathcal{P}\left[  \boldsymbol\eta_{{\boldsymbol \ell},{\boldsymbol s}}{\Phi}^{(-H-1)}_{{\boldsymbol \ell},{\boldsymbol s}}\right](0)\nonumber \\
&+\frac{\sigma}{2\pi} \int_{\mathbb{R}^2} (e^{i{\bf k} \cdot \xx} -1)   \left[ \mathbf{I} - \frac{\kk\kk^T}{ \| {\kk }\|^2}\right]  \boldsymbol\eta_0 \|\kk \|^{-H-1}  \mathcal F{\Phi}^0_0({\bf k})d {\bf k},
\end{align} 
where ${\Phi}^{(-H-1)}_{{\boldsymbol \ell},{\boldsymbol s}}$ and $\mathcal{P}$ are respectively defined in \eqref{eq:fracdiff} and \eqref{eq:ProjLerayDef}, {\color{black} and $\boldsymbol\eta_{{\boldsymbol \ell},{\boldsymbol s}} {\Phi}^{(-H-1)}_{{\boldsymbol \ell},{\boldsymbol s}}$ is the bivariate vector $(\eta^1_{{\boldsymbol \ell},{\boldsymbol s}}{\Phi}^{(-H-1)}_{{\boldsymbol \ell},{\boldsymbol s}},\eta^2_{{\boldsymbol \ell},{\boldsymbol s}}{\Phi}^{(-H-1)}_{{\boldsymbol \ell},{\boldsymbol s}})^T$.}

The decomposition \eqref{eq:WavSeries01} is a consequence of \eqref{wave2}, where $\boldsymbol \epsilon_{{\boldsymbol \ell},{\boldsymbol s}}\triangleq 2\pi\sigma\boldsymbol\eta_{{\boldsymbol \ell},{\boldsymbol s}}$, provided we prove that
\begin{equation}\label{eq:vanish}
\sum_{({\boldsymbol \ell},{\boldsymbol s})\in \Omega} \mathcal{P}\left[  \boldsymbol\eta_{{\boldsymbol \ell},{\boldsymbol s}}{\Phi}^{(-H-1)}_{{\boldsymbol \ell},{\boldsymbol s}}\right](0) =  \frac{1}{(2\pi)^2} \int_{\mathbb{R}^2} (e^{i{\bf k} \cdot \xx} -1)   \left[ \mathbf{I} - \frac{\kk\kk^T}{ \| {\kk }\|^2}\right]  \boldsymbol\eta_0 \|\kk \|^{-H-1}  \mathcal F{\Phi}^0_0({\bf k})d {\bf k}.
\end{equation}

Since by assumption $\int_{[0,1]^2}\tvt(\xx) d\xx=0$, integrating both sides of \eqref{wave2} on $[0,1]^2$ leads to 
\begin{align}\label{eq:int01}
&\sum_{({\boldsymbol \ell},{\boldsymbol s})\in \Omega} \int_{[0,1]^2}\mathcal{P} \left[ \boldsymbol\eta_{{\boldsymbol \ell},{\boldsymbol s}} {\Phi}^{(-H-1)}_{{\boldsymbol \ell},{\boldsymbol s}}\right]({\xx}) d\xx-\mathcal{P}\left[  \boldsymbol\eta_{{\boldsymbol \ell},{\boldsymbol s}}{\Phi}^{(-H-1)}_{{\boldsymbol \ell},{\boldsymbol s}}\right](0) \nonumber\\
&+\frac{1}{(2\pi)^2}\int_{[0,1]^2} d\xx\int_{\mathbb{R}^2} (e^{i{\bf k} \cdot \xx} -1)   \left[ \mathbf{I} - \frac{\kk\kk^T}{ \| {\kk }\|^2}\right]  \boldsymbol\eta_0 \|\kk \|^{-H-1}  \mathcal F{\Phi}^0_0({\bf k})d {\bf k} =0.
\end{align} 
Let \textcolor{black}{ $\boldsymbol z(\xx)= \boldsymbol\eta_{{\boldsymbol \ell},{\boldsymbol s}}{\Phi}^{(-H-1)}_{{\boldsymbol \ell},{\boldsymbol s}}({\xx})$. Since wavelets possess at least one vanishing moment $\int_{[0,1]^2} \boldsymbol z({\xx})d\xx =0$.  According to Definition~\eqref{eq:ProjLerayDef} of the Leray projector this implies that $\int_{[0,1]^2}\mathcal{P}\boldsymbol z(\xx)d\xx=0$ and therefore }
 \begin{align}\label{eq:int01bis}
&\sum_{({\boldsymbol \ell},{\boldsymbol s})\in \Omega} \int_{[0,1]^2}\mathcal{P} \left[ \boldsymbol\eta_{{\boldsymbol \ell},{\boldsymbol s}} {\Phi}^{(-H-1)}_{{\boldsymbol \ell},{\boldsymbol s}}\right]({\xx}) d\xx =0.
\end{align}
Now consider the last term in \eqref{eq:int01}. We have for $j=1,2$
\begin{align*}
&\frac{\partial}{\partial x_j} \int_{\mathbb{R}^2} (e^{i{\bf k} \cdot \xx} -1)    \left[ \mathbf{I} - \frac{\kk\kk^T}{ \| {\kk }\|^2}\right] \boldsymbol\eta_0  \|\kk \|^{-H-1}  \mathcal F{\Phi}^0_0({\bf k})d {\bf k}\\
& =\int_{\mathbb{R}^2} (ik_j)e^{i{\bf k} \cdot \xx}    \left[ \mathbf{I} - \frac{\kk\kk^T}{ \| {\kk }\|^2}\right] \boldsymbol\eta_0  \|\kk \|^{-H-1}  \mathcal F{\Phi}^0_0({\bf k})d {\bf k} \\
&=\int_{\mathbb{R}^2} e^{i{\bf k} \cdot \xx}     \left[ \mathbf{I} - \frac{\kk\kk^T}{ \| {\kk }\|^2}\right] \boldsymbol\eta_0  \|\kk \|^{-H-1}  \left( \int_{\mathbb{R}^2} e^{-i \kk \cdot \xx }\frac{\partial}{\partial x_j} \mathbb{I}_{[0,1]^2}({\xx}) d\xx \right)d {\bf k}=0.
\end{align*}
Therefore, we obtain
 \begin{align}\label{eq:int01tres}
 \int_{[0,1]^2} d\xx\int_{\mathbb{R}^2} (e^{i{\bf k} \cdot \xx} -1)   \left[ \mathbf{I} - \frac{\kk\kk^T}{ \| {\kk }\|^2}\right]  \boldsymbol\eta_0 \|\kk \|^{-H-1}  \mathcal F{\Phi}^0_0({\bf k})d {\bf k}  \nonumber \\
=\int_{\mathbb{R}^2}  (e^{i{\bf k} \cdot \xx} -1)   \left[ \mathbf{I} - \frac{\kk\kk^T}{ \| {\kk }\|^2}\right]  \boldsymbol\eta_0 \|\kk \|^{-H-1}  \mathcal F{\Phi}^0_0({\bf k})d {\bf k}.
 \end{align}
 Using \eqref{eq:int01bis} and \eqref{eq:int01tres} in \eqref{eq:int01} proves \eqref{eq:vanish}, which concludes the proof.

\begin{flushright}{$\square$}\end{flushright}

\subsection{Proof of Proposition~\ref{prop:covDls}}\label{app:proofCovDIs}

{\color{black}
From \eqref{eq:WavSeries01}, we have
 \begin{align*}
d_{{\boldsymbol \ell},{\bf s}}&= \sum_{({\bf i}, {\bf j})\in\Omega}\langle  \mathcal{P} \left[ \boldsymbol\epsilon_ {({\bf i}, {\bf j})} {\Phi}^{(-H-1)}_{({\bf i}, {\bf j})}\right] /  \tilde {\boldsymbol\Psi}_{{\boldsymbol \ell},{\bf s}}\rangle =\sum_{({\bf i}, {\bf j})\in {\Omega}}\langle   \boldsymbol\epsilon_ {({\bf i}, {\bf j})} {\Phi}^{(-H-1)}_{({\bf i}, {\bf j})}/  \mathcal{P} \left[ \tilde {\boldsymbol\Psi}_{{\boldsymbol \ell},{\bf s}}\right]\rangle,
\end{align*}
where the scalar product is in $(L^2([0,1]^2))^2$.  In the above formula and in the following, $\tilde{\boldsymbol\Psi}_{{\boldsymbol \ell},{\bf s}}$ is extended outside $[0,1]^2$ by zero, so that the operation $\mathcal{P} \left[ \tilde {\boldsymbol\Psi}_{{\boldsymbol \ell},{\bf s}}\right]$ makes sense according to Definition~\eqref{eq:ProjLerayDef}. In other words, the definition of $\tilde{\boldsymbol\Psi}_{{\boldsymbol \ell},{\bf s}}$ in Section~\ref{sec:divFreeFracRep}
 becomes   in this case $\mathcal F \tilde{\boldsymbol\Psi}_{{\boldsymbol \ell},{\bf s}} = (- 1/(ik_2) \mathcal F \Phi^0_{{\boldsymbol \ell},{\bf s}}(\kk), (1/ik_1) \mathcal F \Phi^0_{{\boldsymbol \ell},{\bf s}}(\kk))^T$. 

Since $\boldsymbol\epsilon_{({\bf i}, {\bf j})}$ are iid zero-mean Gaussian random variables with variance $(2\pi\sigma)^2\bf I$, we have 
\begin{multline*}
 {(2\pi\sigma)^{-2}} \mathbb{E}[d_{{\boldsymbol \ell},{\bf s}}d_{{\boldsymbol \ell}',{\bf s}'}]=\\
\sum_{({\bf i}, {\bf j})\in\Omega} \langle \Phi^{(-H-1)}_{{\bf i},{\bf j}}, \mathcal{P}^1 \tilde{\boldsymbol\Psi}_{{\boldsymbol \ell},{\bf s}} \rangle  \langle \Phi^{(-H-1)}_{{\bf i},{\bf j}},\mathcal{P}^1  \tilde{\boldsymbol\Psi}_{{\boldsymbol \ell}',{\bf s}'}  \rangle + \langle \Phi^{(-H-1)}_{{\bf i},{\bf j}},\mathcal{P}^2  \tilde{\boldsymbol\Psi}_{{\boldsymbol \ell},{\bf s}} \rangle  \langle \Phi^{(-H-1)}_{{\bf i},{\bf j}},\mathcal{P}^2 \tilde{\boldsymbol\Psi}_{{\boldsymbol \ell}',{\bf s}'} \rangle,
\end{multline*}
where $\mathcal{ P}^k$ is the $k$-th row of matrix operator  $\mathcal{P}$, \ie given in the Fourier domain by $ \mathcal F\mathcal{P}^1= (1-k_1^2/\|\kk\|^2 , -k_1 k_2/\|\kk\|^2)$ and $\mathcal F \mathcal{P}^2= ( -k_1k_2/\|\kk\|^2,1-k_2^2/\|\kk\|^2 )$.

Let us simplify the sum above. First, recall that ${\Phi}^{(-H-1)}_{{\boldsymbol \ell},{\bf s}}(\xx)={(-\Delta)^{\frac{-H-1}{2}}{\Phi}^0_{{\boldsymbol \ell},{\bf s}}}({\xx})$, so for any $k=1,2$: 
\begin{align*}
\sum_{({\bf i}, {\bf j})\in\Omega} \langle \Phi^{(-H-1)}_{{\bf i},{\bf j}},\mathcal{P}^k  \tilde{\boldsymbol\Psi}_{{\boldsymbol \ell},{\bf s}} \rangle & \langle \Phi^{(-H-1)}_{{\bf i},{\bf j}},\mathcal{P}^k  \tilde{\boldsymbol\Psi}_{{\boldsymbol \ell}',{\bf s}'}  \rangle\\ 
&= \langle  \sum_{({\bf i}, {\bf j})\in\Omega}  \Phi^{(-H-1)}_{{\bf i},{\bf j}} \langle \Phi^{(-H-1)}_{{\bf i},{\bf j}},\mathcal{P}^k  \tilde{\boldsymbol\Psi}_{{\boldsymbol \ell}',{\bf s}'}  \rangle ,\mathcal{P}^k \tilde{\boldsymbol\Psi}_{{\boldsymbol \ell},{\bf s}} \rangle  \\
&= \langle  \sum_{({\bf i}, {\bf j})\in\Omega}  \Phi^0_{{\bf i},{\bf j}} \langle \Phi^0_{{\bf i},{\bf j}},(-\Delta)^{\frac{-H-1}{2}}\mathcal{P}^k  \tilde{\boldsymbol\Psi}_{{\boldsymbol \ell}',{\bf s}'}  \rangle ,(-\Delta)^{\frac{-H-1}{2}} \mathcal{P}^k  \tilde{\boldsymbol\Psi}_{{\boldsymbol \ell},{\bf s}}  \rangle.
\end{align*}
Since the mother wavelet $\psi$ has $M>H$ vanishing moments,  similar arguments as  in the proof of Proposition~\ref{corrol01} lead to  $|\|\kk\|^{-H-1}\mathcal F [\mathcal{ P}^k {\tilde{\boldsymbol\Psi}}_{{\boldsymbol \ell'},{\bf s}'}](\kk)|\leq c\|\kk \|^{M-H-1}$, where $c$ is some positive constant.  So  $$\int_{[0,1]^2}(-\Delta)^{\frac{-H-1}{2}}\mathcal{P}^k \tilde{\boldsymbol\Psi}_{{\boldsymbol \ell}',{\bf s}'}(\xx)d\xx=\|\kk\|^{-H-1}\mathcal F [\mathcal{ P}^k {\tilde{\boldsymbol\Psi}}_{{\boldsymbol \ell'},{\bf s}'}](\kk)\Big|_{\kk=0} = 0$$ and we have the equality in  $L^2([0,1]^2)$ : 
\begin{align*}
\sum_{({\bf i}, {\bf j})\in\Omega}  \Phi_{{\bf i},{\bf j}} \langle \Phi^0_{{\bf i},{\bf j}},(-\Delta)^{\frac{-H-1}{2}}\mathcal{P}^k \tilde{\boldsymbol\Psi}_{{\boldsymbol \ell}',{\bf s}'}\rangle&=\sum_{({\bf i}, {\bf j})\in\Omega\cup 0}  \Phi_{{\bf i},{\bf j}} \langle \Phi^0_{{\bf i},{\bf j}},(-\Delta)^{\frac{-H-1}{2}}\mathcal{P}^k\tilde{\boldsymbol\Psi}_{{\boldsymbol \ell}',{\bf s}'}\rangle\\&=(-\Delta)^{\frac{-H-1}{2}}\mathcal{P}^k \tilde{\boldsymbol\Psi}_{{\boldsymbol \ell}',{\bf s}'}.
\end{align*}
Therefore
\begin{align*}
\sum_{({\bf i}, {\bf j})\in\Omega} \langle \Phi^{(-H-1)}_{{\bf i},{\bf j}},\mathcal{P}^k \tilde{\boldsymbol\Psi}_{{\boldsymbol \ell},{\bf s}} \rangle  \langle \Phi^{(-H-1)}_{{\bf i},{\bf j}},\mathcal{P}^k\tilde{\boldsymbol\Psi}_{{\boldsymbol \ell}',{\bf s}'} \rangle &= \langle (-\Delta)^{\frac{-H-1}{2}}\mathcal{P}^k \tilde{\boldsymbol\Psi}_{{\boldsymbol \ell}',{\bf s}'} ,\,(-\Delta)^{\frac{-H-1}{2}} \mathcal{P}^k \tilde{\boldsymbol\Psi}_{{\boldsymbol \ell},{\bf s}}  \rangle
\end{align*}
and 
\begin{equation}\label{cov1}
\mathbb{E}[d_{{\boldsymbol \ell},{\bf s}}d_{{\boldsymbol \ell}',{\bf s}'}]= {(2\pi\sigma)^2}  \langle (-\Delta)^{\frac{-H-1}{2}}\mathcal{P}\tilde{\boldsymbol\Psi}_{{\boldsymbol \ell},{\bf s}} /\, (-\Delta)^{\frac{-H-1}{2}}\mathcal{P}\tilde{\boldsymbol\Psi}_{{\boldsymbol \ell}',{\bf s}'}\rangle.\end{equation}

Since  the operators $\mathcal P$ and $ (-\Delta)^{\frac{-H-1}{2}}$ commute, and $\mathcal P$ is self-adjoint with  $\mathcal P\mathcal P =\mathcal P$, we have 
\begin{equation*}
\mathbb{E}[d_{{\boldsymbol \ell},{\bf s}}d_{{\boldsymbol \ell}',{\bf s}'}]= {(2\pi\sigma)^2}  \langle (-\Delta)^{\frac{-H-1}{2}}\mathcal{P}\tilde{\boldsymbol\Psi}_{{\boldsymbol \ell},{\bf s}} /\, (-\Delta)^{\frac{-H-1}{2}}\tilde{\boldsymbol\Psi}_{{\boldsymbol \ell}',{\bf s}'}\rangle,\end{equation*}
that is by Parseval relation
\begin{align*}
\mathbb{E}[d_{{\boldsymbol \ell},{\bf s}}& d_{{\boldsymbol \ell}',{\bf s}'}]\\
&= {(2\pi\sigma)^2}  \langle \|\kk\|^{-H-1} \left[ \mathbf{I} - \frac{\kk\kk^T}{ \| {\kk }\|^2}\right] \binom{-\frac 1 {ik_2} \mathcal F \Phi^0_{{\boldsymbol \ell},{\bf s}}(\kk)}{\frac 1 {ik_1} \mathcal F \Phi^0_{{\boldsymbol \ell},{\bf s}}(\kk)} /\,  \|\kk\|^{-H-1}  \binom{-\frac 1 {ik_2} \mathcal F \Phi^0_{{\boldsymbol \ell},{\bf s}}(\kk)}{\frac 1 {ik_1} \mathcal F \Phi^0_{{\boldsymbol \ell},{\bf s}}(\kk)}\rangle\\
&={(2\pi\sigma)^2}  \langle \|\kk\|^{-2H-2} \left(\frac{k_2^2}{\|\kk\|^2}\frac 1 {k_2} +\frac{k_1k_2}{\|\kk\|^2} \frac 1 {k_1}\right)\frac{1}{k_2} \mathcal F \Phi^0_{{\boldsymbol \ell},{\bf s}}(\kk),  \mathcal F \Phi^0_{{\boldsymbol \ell},{\bf s}}(\kk)\rangle \\
&\qquad + {(2\pi\sigma)^2}  \langle \|\kk\|^{-2H-2} \left( \frac{k_1k_2}{\|\kk\|^2}\frac 1 {k_2} +\frac{k_1^2}{\|\kk\|^2}\frac 1 {k_1}\right)\frac 1 {k_1} \mathcal F \Phi^0_{{\boldsymbol \ell},{\bf s}}(\kk),  \mathcal F \Phi^0_{{\boldsymbol \ell},{\bf s}}(\kk)\rangle\\
&={(2\pi\sigma)^2}  \langle 4 \|\kk\|^{-2H-4}  \mathcal F \Phi^0_{{\boldsymbol \ell},{\bf s}}(\kk),  \mathcal F \Phi^0_{{\boldsymbol \ell},{\bf s}}(\kk)\rangle.\\
&=4{(2\pi\sigma)^2}  \langle  {\Phi}^{(-H-2)}_{{\boldsymbol \ell},{\bf s}} , {\Phi}^{(-H-2)}_{{\boldsymbol \ell'},{\bf s'}} \rangle,\end{align*}
where the existence of wavelet ${\Phi}^{(-H-2)}_{{\boldsymbol \ell},{\bf s}}$ is guaranteed by the $M>H$ vanishing moments of the mother wavelet $\psi$.
}

\begin{flushright}{$\square$}\end{flushright}

\subsection{Proof of Lemma~\ref{prop:InvCovDls}}\label{appProofDivFreeMap}
{\color{black}
When $n\to\infty$, the matrix ${\bf \Sigma}_n$ becomes  the operator $\underline{\boldsymbol \Sigma}:  \ell^2(\Omega)\rightarrow \ell^2(\Omega)$ defined for any $a\in \ell^2(\Omega)$ by:
\begin{align}\label{eq:Gamma}
[\underline{\boldsymbol \Sigma}a]_{{\boldsymbol \ell},{\boldsymbol s }} &\triangleq \sum_{({\boldsymbol \ell}',{\boldsymbol s }')\in  \Omega} a_{{\boldsymbol \ell}',{\boldsymbol s }'} {\Sigma}({{\boldsymbol \ell},{\boldsymbol s}},{\boldsymbol \ell}',{\boldsymbol s}'), \hspace{0.3cm} \forall ({\boldsymbol \ell},{\boldsymbol s }) \in \Omega,
\end{align}
where ${\Sigma}({{\boldsymbol \ell},{\boldsymbol s}},{\boldsymbol \ell}',{\boldsymbol s}')$ is given by \eqref{defsigma}.
Similarly, the matrix ${\bf \Sigma}_n^{-1}$ becomes  the operator ${\underline{\boldsymbol \Sigma}^{-1}}$ given  for any $a\in \ell^2(\Omega)$ by:
\begin{align}\label{eq:InvGamma}
[\underline{\boldsymbol \Sigma}^{-1}a]_{{\boldsymbol \ell},{\boldsymbol s }}&=\sum_{({\boldsymbol \ell}',{\boldsymbol s }')\in  \Omega} a_{{\boldsymbol \ell}',{\boldsymbol s }'} {\Sigma}^{-1}({{\boldsymbol \ell},{\boldsymbol s}},{\boldsymbol \ell}',{\boldsymbol s}'), \hspace{0.3cm} \forall ({\boldsymbol \ell},{\boldsymbol s }) \in \Omega,
\end{align}
where ${ \Sigma}^{-1}({{\boldsymbol \ell},{\boldsymbol s}},{\boldsymbol \ell},{\boldsymbol s})$ is given by \eqref{definvsigma}.

We denote  $ \Phi^{(-H-2)}_{per;\boldsymbol \ell,{\boldsymbol s}} \triangleq \mathcal F_{per}^{-1}\left( \left\{ \| {\bf k } \| ^{-H-2}\mathcal F_{per} ( {\Phi}_{{\boldsymbol \ell},{\boldsymbol s}})(\kk)\right\}_{\kk\in\mathbb Z^2}\right)$ and similarly $\Phi^{(H+2)}_{per;\boldsymbol \ell,{\boldsymbol s}}$, that are well-defined since the mother wavelet $\psi$ has $M>H$ vanishing moments and is $H+2$ times differentiable. Using the fact that the wavelets $\Phi^{(-H-2)}_{per;{\boldsymbol \ell},{\boldsymbol s}}$  and the dual wavelets $\Phi^{(H+2)}_{per;{\boldsymbol \ell},{\boldsymbol s}}$ form a biorthogonal basis of $L^2([0,1]^2)$ when ${({\boldsymbol \ell},{\boldsymbol s})} \in \Omega$,
we have for any $a\in\ell^2(\Omega)$: 
\begin{align*}
[\underline{\boldsymbol\Sigma}\underline{\boldsymbol\Sigma}^{-1}a]_{{\boldsymbol \ell},{\boldsymbol s}} &=\sum_{({\bf i}, {\bf j})\in\Omega} \sum_{({\boldsymbol \ell}',{\boldsymbol s}')\in\Omega}a_{\boldsymbol \ell',{\boldsymbol s}'}  { \Sigma}^{-1}({{\bf i},{\bf  j}},{\boldsymbol \ell}',{\boldsymbol s}'){\Sigma}({{\boldsymbol \ell},{\boldsymbol s}},{\bf i},{\bf j})\\
&= \sum_{({\boldsymbol \ell}',{\boldsymbol s}')\in\Omega}\sum_{({\bf i}, {\bf j})\in\Omega}a_{\boldsymbol \ell',{\boldsymbol s}'}\langle \Phi^{(H+2)}_{per;\boldsymbol \ell',{\boldsymbol s}'},\Phi^{(H+2)}_{per;{\bf i}, {\bf j}}\rangle\langle \Phi^{(-H-2)}_{per;\boldsymbol \ell,{\boldsymbol s}},\Phi^{(-H-2)}_{per;{\bf i}, {\bf j}}\rangle\\
&=\sum_{({\boldsymbol \ell}',{\boldsymbol s}')\in\Omega} a_{\boldsymbol \ell',{\boldsymbol s}'} \langle  \Phi^{(H+2)}_{per;\boldsymbol \ell',{\boldsymbol s}'},\sum_{({\bf i}, {\bf j})\in\Omega} \Phi^{(H+2)}_{per;{\bf i}, {\bf j}} \langle \Phi^{(-H-2)}_{per;\boldsymbol \ell,{\boldsymbol s}},\Phi^{(-H-2)}_{per;{\bf i}, {\bf j}}\rangle\rangle\\
&=\sum_{({\boldsymbol \ell}',{\boldsymbol s}')\in\Omega} a_{\boldsymbol \ell',{\boldsymbol s}'} \langle  \Phi^{(H+2)}_{per;\boldsymbol \ell',{\boldsymbol s}'}, \Phi^{(-H-2)}_{per;\boldsymbol \ell,{\boldsymbol s}}\rangle\\
&=a_{\boldsymbol \ell,{\boldsymbol s}}.
\end{align*}
We can show similarly that $\underline{\boldsymbol\Sigma}^{-1}\underline{\boldsymbol\Sigma}a=a$. Therefore  operator $\underline{\boldsymbol\Sigma}^{-1}$ corresponds to the inverse operator of $\underline{\boldsymbol\Sigma}$.
}

 \begin{flushright}{$\square$}\end{flushright}

\subsection{Proof of Proposition~\ref{lemma:specGrad}}\label{app:specGrad}

The gradient with respect to ${\epsilon^1_{{\boldsymbol \ell},{\boldsymbol s}}}$ of the data-term $\delta y({\boldsymbol \epsilon_n})$  in \eqref{objFunction6} is given by inner products with the fractional divergence-free wavelets. Indeed,   we have:
\begin{align*}
\partial_{\epsilon^1_{{\boldsymbol \ell},{\boldsymbol s}}} \delta y({\boldsymbol {\boldsymbol \epsilon_n}}) &=  \langle ( \bar y_1(\xx,{\boldsymbol \epsilon_n})-y_0(\xx)) ,  \nabla    \bar y_1(\xx,{\boldsymbol \epsilon_n})^T \frac{\partial }{\partial {\epsilon^1_{{\boldsymbol \ell},{\boldsymbol s}}} }{\color{black} \mathcal{P}_{per} }\left[ {\underline{\boldsymbol \Phi}^{(-H-1)}_{  n}} {\boldsymbol \epsilon}_n\right] (\xx)  \rangle \nonumber \\
&= \langle  \bar y_1(\xx,{\boldsymbol \epsilon_n})-y_0(\xx))   \nabla    \bar y_1(\xx,{\boldsymbol \epsilon_n}) /  \frac{\partial }{\partial {\epsilon^1_{{\boldsymbol \ell},{\boldsymbol s}}} }{\color{black} \mathcal{P}_{per}} \left[  \boldsymbol\epsilon_{{\boldsymbol \ell},{\boldsymbol s}}{\Phi}^{(-H-1)}_{{\boldsymbol \ell},{\boldsymbol s}}\right] ({\xx})  \rangle \nonumber
 \end{align*}
and by Fourier-Plancherel formula:
{\color{black}
\begin{align*}
& \partial_{\epsilon^1_{{\boldsymbol \ell},{\boldsymbol s}}} \delta y({\boldsymbol \epsilon_n})= \\
&  \langle\mathcal F_{per}\left((\bar y_1(\kk,{\boldsymbol \epsilon_n}) -y_0(\kk) )\nabla   \bar  y_1(\kk,{\boldsymbol \epsilon_n})\right) /  \frac{\partial }{ \partial {\epsilon^1_{{\boldsymbol \ell},{\boldsymbol s}}} }   \left[ \mathbf{I} - \frac{\kk\kk^T}{ \| {\kk }\|^2}\right]  \boldsymbol\epsilon_{{\boldsymbol \ell},{\boldsymbol s}} \|\kk\|^{\!-\!H\!-\!1\!}\mathcal F_{per}({\Phi}_{{\boldsymbol \ell},{\boldsymbol s}})({\kk})   \rangle \nonumber \\
 &=  \langle\mathcal F_{per}\left((\bar y_1(\kk,{\boldsymbol \epsilon_n}) -y_0(\kk) )\nabla   \bar  y_1(\kk,{\boldsymbol \epsilon_n})\right) / \|\kk\|^{\!-\!H\!-\!1\!} \begin{pmatrix} 1- \frac{k_1^2}{\|k\|^2}\\ -\frac{k_1k_2}{\|k\|^2}\end{pmatrix}  \mathcal F_{per}({\Phi}_{{\boldsymbol \ell},{\boldsymbol s}})({\kk})  \rangle \nonumber \\
&=\langle \|\kk\|^{\!-\!H\!-\!1\!}   \begin{pmatrix} 1- \frac{k_1^2}{\|k\|^2}\\ -\frac{k_1k_2}{\|k\|^2}\end{pmatrix}^T\mathcal F_{per}\left((\bar y_1(\kk,{\boldsymbol \epsilon_n}) -y_0(\kk) )\nabla   \bar  y_1(\kk,{\boldsymbol \epsilon_n})\right), \mathcal F_{per}( {\Phi}_{{\boldsymbol \ell},{\boldsymbol s}}) (\kk)\rangle 
 \end{align*}
where scalar products are in $\ell^2(\mathbb{Z}^2)$. Hence \eqref{gradAdjoint} is deduced when $i=1$ by Parseval formula. 
} 
 The gradient with respect to ${\epsilon^2_{{\boldsymbol \ell},{\boldsymbol s}}}$ is obtain similarly.
The gradient of the  regularization term is simply: 
$\partial_{{\boldsymbol \epsilon_n}} \mathcal{R}({\boldsymbol \epsilon_n})={\color{black}\frac{1}{\beta\sigma^2}}{\boldsymbol \epsilon_n}.$ {$\square$}

\subsection{Proof of Proposition~\ref{prop:noApproxDiv0Prior}}\label{app:noApproxDiv0Prior}

The gradient \eqref{gradJlDivFree}  of the data-term $\delta y( d_n)$ is obtained  analogously to \eqref{gradAdjoint}.
For the gradient of the regularizer term, by Definition~\eqref{definvsigma} of ${\boldsymbol \Sigma}^{-1}_n$:
 \begin{align*}
\partial_{d_{{\boldsymbol \ell},{\boldsymbol s}}} \mathcal{R}({ d_n})&=\frac{1}\beta \sum_{({{\boldsymbol \ell}',{\boldsymbol s}'}) \in \Omega_n} {\boldsymbol \Sigma}^{-1}_n({\boldsymbol \ell},{\boldsymbol s},{\boldsymbol \ell}',{\boldsymbol s}') d_{{\boldsymbol \ell}',{\boldsymbol s}'}\\
&={\color{black} \frac{1}{\beta\sigma^2}} \sum_{({{\boldsymbol \ell}',{\boldsymbol s}'}) \in \Omega_n}  \langle  {\Phi}^{(H+2)}_{{\boldsymbol \ell},{\bf s}} , {\Phi}^{(H+2)}_{{\boldsymbol \ell'},{\bf s'}} \rangle d_{{\boldsymbol \ell}',{\boldsymbol s}'}
\end{align*}
Formula \eqref{eq:gradfuncDiv0Noapprox} follows from definition of  ${\underline{\boldsymbol \Phi}^{1,{(H+2)}}_{  n}}$, see \eqref{eq:operatorFracNewDef}.

  \begin{flushright}{$\square$}\end{flushright}

%
%
%

\section{Adaptation of algorithm \ref{algo5} for irregular wavelets}\label{app:Algo5Modif}

 If $\psi$ is not $2H+4$ but only {\color{black}$\max(2H,H+2)$} times differentiable, one may replace  $iii)$-$v)$ in algorithm \ref{algo5} by the following steps:

\begin{itemize} 
\item[-]
  compute {\color{black} $\mathcal{F}^{-1}_{per}  \left( \| \kk\|^{2H} \mathcal{F}_{per}({\underline{\boldsymbol \Phi}_{  n}^{1,(0)}} d_n )\right) $}  by FFT
 \item[-] compute  the FWT  using  orthogonal wavelets $\{\Phi_{{\boldsymbol \ell},{\boldsymbol s}};({{\boldsymbol \ell},{\boldsymbol s}})\in \Omega_n\}$ to get 
the $n \times n$ matrix denoted by $\llbracket { e_n}\rrbracket$   whose element at  row index $({{ \ell}_1,{ s}_1})$ and column index $({{ \ell}_2,{ s}_2} ) $ is {\color{black}  $\langle \mathcal{F}^{-1}_{per} \left( \| \kk\|^{2H} \mathcal{F}_{per}({\underline{\boldsymbol \Phi}_{  n}^{1,(0)}} d_n ) \right), \Phi_{{\boldsymbol \ell},{\boldsymbol s}} \rangle$. }
 \item[-] compute off-line matrices $\FF^ {(\alpha)}$ for $\alpha=0,1,2$ with algorithm \ref{algo6} (section \ref{sec:SpatialOptim})
  \item[-] obtain the scalar product in \eqref{eq:gradfuncDiv0Noapprox} by addition of  matrix products 
  \begin{align*}
  \langle  {\underline{\boldsymbol \Phi}^{1,{(H+2)}}_{  n}} & d_n ,  {\Phi}^{(H+2)}_{{\boldsymbol \ell},{\boldsymbol s}} \rangle  \\
  & =\sum_{({\boldsymbol \ell}',{\boldsymbol s }')\in  \Omega_n} \bigg( \langle  \mathcal{F}^{-1}_{per}  \left( \| \kk\|^{2H} \mathcal{F}_{per}({\underline{\boldsymbol \Phi}_{  n}^{1,(0)}} d_n ) \right), \Phi_{{\boldsymbol \ell}',{\boldsymbol s}'} \rangle \\   
  & \quad\qquad \langle \mathcal{F}^{-1}_{per}  \left( \| \kk\|^2 \mathcal{F}_{per}( \Phi_{{\boldsymbol \ell}',{\boldsymbol s}'})\right),  \mathcal{F}^{-1}_{per}  \left( \| \kk\|^2 \mathcal{F}_{per}( \Phi_{{\boldsymbol \ell},{\boldsymbol s}})\right)  \rangle\bigg) \\
  &=\left( \FF^{(2)^T}\llbracket { e_n}   \rrbracket \FF^ {(0)} + 2\FF^{(1)^T}\llbracket { e_n}   \rrbracket \FF^ {(1)} +\FF^{(0)^T}\llbracket { e_n}   \rrbracket \FF^ {(2)} \right)_{{\boldsymbol \ell},{\boldsymbol s}},
  \end{align*}
  where $({\mathbf M})_{{\boldsymbol \ell},{\boldsymbol s}}$ denotes the $({{\boldsymbol \ell},{\boldsymbol s}})$-th element  of matrix ${\mathbf M}$
  \end{itemize}

\section{Matrices of  mono-dimensional connection coefficients}\label{app1}

 The matrices $\FF^{(\alpha)}$ involved in  \eqref{eq:lowDimMatProdGrad}, where $0\leq \alpha<H+2$, are composed of wavelets connection coefficients defined in \eqref{eq:defCoeffConnec}. Note that  $ \FF^ {(0)}$ is the identity matrix since we are considering an orthonormal basis. Moreover, for $\alpha$ being a positive integer,  fractional Laplacian operator {\color{black} $\mathbf{v} \mapsto \mathcal{F}^{-1}_{per}  \left( \| \kk\|^{\alpha} \mathcal{F}_{per}(\mathbf{v})\right)$} becomes a standard differentiation up to factor $(-2\pi)^{\alpha}$ and $\FF^{(\alpha)}$ can be computed by solving an eigenvalue problem as detailed in  \cite{Beylkin92,Dahmen93}.  However, in the more general case of fractional Laplacian differentiation, no method have been explicitly proposed  in literature.   In this appendix we  provide an approximation of   $\FF^{(\alpha)}$ in terms of {\it scaling functions}  connection coefficients,  that turn out  to be easily computable as the solution of a linear system, as explained in the following.

 \subsection{Matrix $\FF^{(\alpha)}$}\label{app1:sub1}
 In this section we  assume $\alpha>0$ and we show that  any entry $f^{(\alpha)}_{\ell,s,\ell',s'} $ of $\FF^{(\alpha)}$ can be determined recursively from an infinite series of connection coefficients of scaling functions defined at the finest  scale $s_n=log_2(n)$. These connection coefficients, denoted by $e_{s_n}^{(\alpha)}$, are given for any $\ell$, $\ell' \in \mathbb{Z}$ by 
\begin{align}  \label{eq:ScalConnec}
 e_{s_n}^{(\alpha)}(\ell,\ell') & \triangleq  \langle \varphi(2^{s_n}x-\ell), \left( \frac{-\partial^2}{\partial x^2}\right)^{\alpha}\varphi(2^{s_n}x-\ell' )\rangle_{\mathbb{R}}  
\end{align}
 where $\varphi$ denotes the scaling function associated to wavelet $\psi$. An efficient algorithm for the  computation of   $e_{s_n}^{(\alpha)}$ is obtained in section \ref{app1:sub2}. We hereafter propose an approximation of  $   f^{(\alpha)}_{\ell,s,\ell',s'} $ as a truncation of these infinite series of scaling function connection coefficients. 
  
Let us begin by recalling the two-scale relations associated to the orthonormal wavelet basis  $\{\psi_{\ell,s}(x); x\in\mathbb R, \ell, s \in \mathbb{Z}\}$ defined in \eqref{defpsi}, see \cite{mallat2008wavelet}:
  \begin{align}\label{eq:2scaleRelations}
  \varphi(x)=\sqrt{2} \sum_k h_k\varphi(2x-k),\\
  \psi(x)=\sqrt{2} \sum_k g_k\varphi(2x-k),
  \end{align}  
where $h$ and $g$ are the conjugate mirror filters of finite impulse response given by $h_k= \frac{1}{\sqrt{2}} \langle  \varphi(x), \varphi(2x-k)\rangle$ and $g_k= \frac{1}{\sqrt{2}} \langle    \psi(x), \varphi(2x-k)\rangle$.  
 For any function $b(\ell,\ell') \in L^2(\mathbb{Z}^2) $, let us define the following  convolution operators:
\begin{align}
\underline{G}_1 b(\ell,\ell') &\triangleq \sum_k g_k b(2\ell+k,\ell').\label{eq:OpFWTDef0}\\
\underline{G}_2b (\ell,\ell') &\triangleq \sum_k g_k b(\ell,2\ell'+k) \\
\underline{H}_1 b (\ell,\ell') &\triangleq \sqrt{2}\sum_k h_k b(2\ell+k,\ell') \\
\underline{H}_2  b (\ell,\ell') &\triangleq \sqrt{2}\sum_k h_kb(\ell,2\ell'+k).\label{eq:OpFWTDef}
\end{align}
We also consider operator  $\underline{H}_1^{(i)}$ (resp. $\underline{H}_2^{(i)}$)  defined by iterating $i$  times  operator  $\underline{H}_1$ (resp.   $\underline{H}_2$).
  Following the methodology introduced in \cite{Beylkin91} (see details in \cite{Perrier99}),  we obtain from \eqref{eq:2scaleRelations}-\eqref{eq:OpFWTDef} that  
   for  $ { s}, s'  \le s_n$  
\begin{align} \label{eq:FWTscalingFunc}
  \langle\psi_{{ \ell},s}, {\left(\frac{-\partial^2}{\partial x^2}\right)}^{\alpha}{\psi}_{\ell',s'} \rangle &= \underline{G}_1\underline{G}_2 \underline{H}_1^{(s_n-s)}\underline{H}_2^{(s_n-s')}  e_{s_n}^ {(\alpha)} (\ell,\ell').
  \end{align}
  
To get a similar representation for  $f^{(\alpha)}_{\ell,s,\ell',s'}$, we need to consider the same procedure with periodized wavelets and scaling functions instead of $\psi$ and $\varphi$. It can be shown that in the case of scaling functions defined at scale ${s_n}$ and periodized over $[0,1]$,  connection coefficients  are  $(2^{s_n})$-periodic functions 
  \begin{align*}
 \sum_{k,k'=-\infty}^{+\infty} \langle  \varphi(2^{s_n}x-\ell+k) , \left( \frac{-\partial^2}{\partial x^2}\right)^{\alpha} \varphi(2^{s_n}x-\ell' +k')\rangle_{[0,1]}   =\sum_{k=-\infty}^{+\infty}  e_{s_n}^{(\alpha)}(\ell+k2^{s_n},\ell'),
  \end{align*}
 provided the latter series converges.
Therefore,  by redefining operators \eqref{eq:OpFWTDef0}-\eqref{eq:OpFWTDef} with circular convolution on $(2^{s_n})$-periodic signals,  we obtain similarly  as \eqref{eq:FWTscalingFunc}: for  $0< { s}, s'  \le s_n$ and for $ 0\le\ell \!<\!2^{s-1}, 0\le \ell'\!<\!2^{s'-1}$, 
 \begin{align} \label{eq:FWTscalingFunc3}
   f^{(\alpha)}_{\ell,s,\ell',s'} 
   &={\color{black} (2\pi)^{-2\alpha}}\underline{G}_1\underline{G}_2 \underline{H}_1^{(s_n-s)}\underline{H}_2^{(s_n-s')}\sum_{k=-\infty}^{+\infty}  e_{s_n}^{(\alpha)}(\ell+k2^{s_n},\ell'),
  \end{align}
  provided the latter series is convergent.
The remaining terms of $\FF^{(\alpha)}$ can  be treated in the same way: for $0< s  \le s_n,\, 0\le  \ell\!<\!2^{s-1}$
 \begin{equation} \label{eq:FWTscalingFunc2}
\left\{\begin{aligned}
   f^{(\alpha)}_{0,0,\ell,s} 
   &={\color{black} (2\pi)^{-2\alpha}}\underline{G}_2 \underline{H}_1^{(s_n)}\underline{H}_2^{(s_n-s)}\sum_{k=-\infty}^{+\infty}  e_{s_n}^{(\alpha)}(k2^{s_n},\ell)\\
  f^{(\alpha)}_{\ell,s,0,0} 
   &={\color{black} (2\pi)^{-2\alpha}}\underline{G}_1 \underline{H}_1^{(s_n-s)}\underline{H}_2^{(s_n)}\sum_{k=-\infty}^{+\infty}  e_{s_n}^{(\alpha)}(\ell+k2^{s_n},0)\\
 f^{(\alpha)}_{0,0,0,0} 
   &= {\color{black} (2\pi)^{-2\alpha}}\underline{H}_1^{(s_n)}\underline{H}_2^{(s_n)}\sum_{k=-\infty}^{+\infty}  e_{s_n}^{(\alpha)}(k2^{s_n},0)
\end{aligned}\right..
 \end{equation}
Recursive formulae \eqref{eq:FWTscalingFunc3}-\eqref{eq:FWTscalingFunc2} show that  the knowledge of  $e_{s_n}^{(\alpha)}$ entirely determines  the matrix $\FF^{\alpha}$. 

Finally, as it will be explained in section  \ref{app1:sub2}, $e_{s_n}^{(\alpha)}(\ell,0)\sim c_\alpha |\ell|^{-1-2\alpha}$ as $\ell\to\infty$, where $c_\alpha>0$.  Since $e_{s_n}^{(\alpha)}(\ell,\ell')=e_{s_n}^{(\alpha)}(\ell-\ell',0)$, we deduce that for any $0\le \ell, \ell'\!<\!2^{s_n}$, and for any $k\neq 0$,  $e_{s_n}^{(\alpha)}(\ell+k2^{s_n},\ell')$ behaves as $c_\alpha |\ell'-\ell-k 2^{s_n}|^{-1-2\alpha}$ if $2^{s_n}$ is sufficiently large. This shows that the terms associated to $k\neq 0$ in  \eqref{eq:FWTscalingFunc3}-\eqref{eq:FWTscalingFunc2} are negligible with respect to the the terms associated to $k=0$, provided $2^{s_n}$ is sufficiently large. The latter condition is a reasonable assumption in standard image setting  where typically  $s_n\geq 8$. This is the reason why we propose the following approximation, for any $0\le \ell, \ell'\!<\!2^{s_n}$:
\begin{equation}\label{eq:approxAsymptoConnec}
\sum_{k=-\infty}^{+\infty}  e_{s_n}^{(\alpha)}(\ell+k2^{s_n},\ell')  \approx
 \begin{cases}
 e_{s_n}^{(\alpha)}(\ell,\ell') & \textrm{for} \hspace{0.3cm}0\le  |\ell' -\ell| < 2^{s_n-1},\\
 e_{s_n}^{(\alpha)}(\ell,\ell'-2^{s_n}) & \textrm{for} \hspace{0.3cm}2^{s_n-1} \le |\ell' -\ell| < 2^{s_n}.
\end{cases}
\end{equation}
This approximation is based on the above explanation when $|\ell' -\ell| < 2^{s_n-1}$ and is extended to $2^{s_n-1} \le |\ell' -\ell| < 2^{s_n}$ in order to respect  $(2^{s_n})$-periodicity.

The derivation of  matrix $\FF^{(\alpha)}$ is thus very simple:  from \eqref{eq:FWTscalingFunc3}-\eqref{eq:FWTscalingFunc2}, we see that matrix $\FF^{(\alpha)}$ is a   bi-dimensional anisotropic  discrete wavelet transform of the   $(2^{s_n})$-periodic function $\sum_{k=-\infty}^{+\infty}  e_{s_n}^{(\alpha)}(\ell+k2^{s_n},\ell')$, where the latter is approximated by  \eqref{eq:approxAsymptoConnec}.  In other words, relations \eqref{eq:FWTscalingFunc3}-\eqref{eq:FWTscalingFunc2}  perform a basis change from the  orthonormal family  $\{\sum_{k,k'}\varphi(2^{s_n}(x_1+k)-\ell)\varphi(2^{s_n}(x_2+k)-\ell'); 0\le \ell,\ell' < 2^{s_n}\}$ to the orthonormal   family $\{{\Phi}_{{\boldsymbol \ell},{\boldsymbol s}}; ({{\boldsymbol \ell},{\boldsymbol s}}) \in \Omega_n\cup 0\}$.
 Indeed,  recursive  convolutions appearing in \eqref{eq:FWTscalingFunc3}-\eqref{eq:FWTscalingFunc2}  implement  (up to the multiplicative constant $2^{s_n}$) the fast recursive filtering algorithm proposed by Mallat \cite{mallat2008wavelet} for FWT.  In practice, we thus  compute   $\FF^{(\alpha)}$ by a simple  FWT of  the discrete function defined in the right hand side of\eqref{eq:approxAsymptoConnec} {\color{black}multiplied by factor $ (2\pi)^{-2\alpha}$}.

\subsection{Computation of connection coefficients  $e_{s_n}^{(\alpha)}$}\label{app1:sub2}

 We  hereafter adapt the   general framework proposed by Beylkin in  {\cite{Beylkin92,Beylkin93}} to the case of the computation of scaling function connection coefficients appearing  in \eqref{eq:approxAsymptoConnec}.

The fractional Laplacian operator   is rewritten as a convolution operator for any scaling function with a compact support. Indeed, if  $\alpha-1/2 \in \mathbb{R}\backslash \mathbb{N}$, fractional Laplacian can also be defined by Riesz potential\footnote{This definition can be extend to $\alpha-1/2 \in \mathbb{N}$ using some appropriate kernel, see \eg \cite{Reichel09}} \cite{Gorenflo98}: 
\begin{align*}
\varphi^{(\alpha)}(x)\triangleq \left( \frac{-\partial^2}{\partial x^2}\right)^{{\alpha}} \varphi(x)= \frac{1}{c_\alpha}\int_{-\infty}^{+\infty} \varphi(z)\frac{1}{|x-z|^{2\alpha+1}}dz,
\end{align*}
with $ c_\alpha=\frac{\sqrt{\pi}\Gamma(-\alpha)2^{-2\alpha}} {\Gamma((1+2\alpha)/2)}$. In the previous expression, the convolution kernel writes 
\begin{align}\label{eq:kernel}
k(x)=\frac{1}{c_\alpha |x|^{2\alpha+1}}.
\end{align}
 
 Since we have $e_{s_n}^{(\alpha)}(\ell,\ell')= e_{s_n}^{(\alpha)}(\ell-\ell',0)$, the computation of all scaling function connection coefficients reduces to the computation of $ e_{s_n}^{(\alpha)}(\ell,0)$ for $\ell\in\mathbb Z$.
  From   \eqref{eq:2scaleRelations}, we derive that  
 \begin{align*}
 \varphi^{(\alpha)}(x)&=\sqrt{2}\sum_{k=0}^{L-1}h_k \left(\frac{-\partial^2}{\partial x^2}\right)^{{\alpha}}\varphi(2x-k) =\sqrt{2}2^{2\alpha}\sum_{k=0}^{L-1}h_k \varphi^{(\alpha)}(2x-k),
  \end{align*}
 where  $L$ denotes the number of non zero coefficients of the scaling  filter $h$. Using    \eqref{eq:2scaleRelations} for $\varphi$ and the above relation for   $\varphi^{(\alpha)}$ in  \eqref{eq:ScalConnec} leads to  
\begin{align}\label{recursion}
 e_{s_n}^{(\alpha)}(\ell,0) =2^{2\alpha} \sum_{k=0}^{L-1} \sum_{j=2\ell-k}^{L-1+2\ell-k} h_k h_{k-2\ell+j}\, e_{s_n}^{(\alpha)}(j,0).
\end{align} 

Moreover,  an asymptotic behavior can be derived from the Taylor expansion of the kernel \eqref{eq:kernel} as in \cite{Beylkin92,Beylkin93}:  for   $\ell\to\infty$, 
\begin{align}\label{e_asymptotic}
 e_{s_n}^{(\alpha)}(\ell,0)&= \frac{1}{c_\alpha |\ell|^{1+2\alpha}} + \mathcal{O}\left(\frac{1}{|\ell|^{1+2\alpha+2M}} \right). \end{align} 

In order to compute $e_{s_n}^{(\alpha)}(\ell,0)$, we solve the linear system \eqref{recursion} subjected to the above asymptotic behavior as boundary conditions.
Specifically, for $|\ell|>\ell_{\textrm{min}}$, where  $\ell_{\textrm{min}}$ is chosen sufficiently large (typically $\ell_{\textrm{min}}>n/8$), we set $e_{s_n}^{(\alpha)}(\ell,0)= \frac{1}{c_\alpha |\ell|^{1+2\alpha}}$. Then for $\ell=-\ell_{\textrm{min}},...,\ell_{\textrm{min}}$, an analytical solution $ {e}^{({\alpha})}_{s_n}(\ell,0)$ of  \eqref{recursion} is obtained as described below. 

Let ${ H}_{k}$ be the function  defined for any of $\ell, j \in \mathbb{Z}$ by 
\begin{align*}
{ H}_{k}({\ell,j})=\begin{cases} h_kh_{k-2\ell+j} &  \textrm{if} \hspace{0.3cm} 2\ell -k \le j\le L-1+ 2\ell -k,  \\
0 &\textrm{otherwise}.\end{cases}
\end{align*}
Let ${\bf H}_{k}$ be the matrix of size $(2 \ell_{min}+1) \times  (2 \ell_{min}+1)$ whose element at row $\ell$ and column $j$ is  ${ H}_{k}({\ell-\ell_{min},j-\ell_{min}})$. Let   ${\bf I}$ denote the  identity matrix. Then 
\begin{align}\label{eq:invSystem}
 {\bf e}^{({\alpha})}_{s_n}&= [2^{2\alpha}\sum_{k=0}^{L-1}{\bf H}_{k}-{\bf I}]^{-1}{\bf b}
\end{align}
where $ {\bf e}^{({\alpha})}_{s_n}$ and ${\bf b}$ are $(2\ell_{\textrm{min}}+1)$-dimensional  vector whose components    
are       respectively $ {e}^{({\alpha})}_{s_n}(\ell,0)$ and $${\bf b}_{\ell}=-2^{2\alpha}\left(\sum_{j=\ell_{\textrm{min}}+1}^{\ell_{\textrm{min}}+L}\sum_{k=0}^{L-1}\frac{ {H}_{k}(\ell-\ell_{min},j)}{c_\alpha |j|^{1+2\alpha}} + \sum_{j=-\ell_{\textrm{min}}-L}^{-\ell_{\textrm{min}}-1}\sum_{k=0}^{L-1}\frac{ {H}_{k}(\ell-\ell_{min},j)}{c_\alpha |j|^{1+2\alpha}}\right).
$$

 \bibliographystyle{spmpsci}      
\bibliography{fluminance}

{\footnotesize 
\noindent
\begin{figure}	[!ht]
\hspace{-0.cm}\begin{tabular}{|c|c|c|c|c|c|c|}
\hline
&\multicolumn{4}{c}{}&\\
$ H$&\multicolumn{4}{c}{{\bf RMSE/MBAE/SAE}}&\\
&\multicolumn{4}{c}{}&\\
\hline
&{\bf A} &{\bf B }&{\bf C} & {\bf D}   &{\bf E}\\
\hline
 0.01& 2.03/37.67/10.12&2.09/38.59/12.95&\textcolor{blue}{1.69}/\textcolor{blue}{30.23}/\textcolor{blue}{2.34}&\textcolor{blue}{1.96}/\textcolor{blue}{34.46}/\textcolor{blue}{1.88}&\textcolor{blue}{1.96}/\textcolor{blue}{35.73}/\textcolor{blue}{2.08}\\
\hline
 $1/3$& 1.50/19.60/5.08&1.55/20.44/11.55&\textcolor{blue}{1.15}/\textcolor{blue}{15.18}/\textcolor{blue}{3.20}&\textcolor{blue}{1.35}/\textcolor{blue}{17.41}/\textcolor{blue}{1.01}&\textcolor{blue}{1.36}/\textcolor{blue}{17.52}/\textcolor{blue}{1.11}\\
\hline
 $1/2$&1.19/12.71/4.04&1.25/13.55/11.10&\textcolor{blue}{0.88}/\textcolor{blue}{9.52}/\textcolor{blue}{3.17}&\textcolor{blue}{1.14}/\textcolor{blue}{12.31}/\textcolor{blue}{1.40}&\textcolor{blue}{1.11}/\textcolor{blue}{11.98}/\textcolor{blue}{1.31} \\
\hline
$2/3$ & 0.89/7.77/\textcolor{blue}{4.16}&0.93/8.37/10.67&\textcolor{blue}{0.68}/\textcolor{blue}{6.08}/10.16 &\textcolor{blue}{0.87}/\textcolor{blue}{7.55}/\textcolor{blue}{0.84}&\textcolor{blue}{0.85}/\textcolor{blue}{7.51}/\textcolor{blue}{1.00}\\
\hline
1& 0.46/3.40/\textcolor{blue}{3.25}&0.45/3.38/9.87&\textcolor{blue}{ 0.41}/\textcolor{blue}{3.21}/9.18&\textcolor{blue}{0.44}/\textcolor{blue}{3.27}/\textcolor{blue}{2.28}&  \textcolor{blue}{0.43}/\textcolor{blue}{3.24}/\textcolor{blue}{2.23} \\
\hline
\end{tabular}	\caption{{\footnotesize Performance of the regularizers according to the value of $H$. Regularization coefficient  for methods A to E (see section \ref{sec:refmethods} for a description) were chosen to  minimize the RMSE.  The given criteria are in order RMSE/MBAE/SAE. For each value of $H$, the 3 best results with respect to each criterion are displayed in blue.\label{fig:performance}}}\vspace{-0.cm}
\end{figure}
}

%
%
%

\end{document}